\documentclass[12pt,epsf]{article}
\usepackage{amssymb,amsmath,amsbsy}
\usepackage{graphicx,color}
\definecolor{purple}{rgb}{0.7,0.0,0.5}
\usepackage[bookmarks = false, colorlinks = true, linkcolor = black, citecolor = black, urlcolor=black]{hyperref}
\newcommand{\be}{\begin{equation}}
\newcommand{\ee}{\end{equation}}
\newcommand{\bea}{\begin{eqnarray}}
\newcommand{\eea}{\end{eqnarray}}
\newcommand{\bear}{\begin{eqnarray}}
\newcommand{\eear}{\end{eqnarray}}
\newcommand{\beas}{\begin{eqnarray*}}
\newcommand{\p}{\partial}
\newcommand{\ep}{\epsilon}
\newcommand{\eeas}{\end{eqnarray*}}
\newcommand{\ba}{\begin{array}}
\newcommand{\ea}{\end{array}}

\newcommand{\nn}{\nonumber}
\newcommand{\lb}{\left(}
\newcommand{\rb}{\right)}
\newcommand{\la}{\langle}
\newcommand{\ra}{\rangle}

\newcommand{\tr}{\operatorname{tr}}
\newcommand{\pd}[2][1]{\ifnum#1=1 \frac{\partial}{\partial {#2}} \else
  \frac{\partial^#1}{\partial {#2}^{#1}}\fi}
\newcommand{\dpd}[2][1]{\ifnum#1=1 \dfrac{\partial}{\partial {#2}} \else
  \frac{\partial^#1}{\partial {#2}^{#1}}\fi}
\newcommand{\td}[2][1]{\ifnum#1=1 \frac{d}{d{#2}} \else
  \frac{d^#1}{d{#2}^{#1}}\fi}

\newcommand{\KK}{\mathcal{K}}
\newcommand{\LL}{\mathcal{L}}
\newcommand{\MM}{\mathcal{M}}

\newcommand{\OO}{\mathcal{O}}

\newcommand{\lsb}{\left[}
\newcommand{\rsb}{\right]}

\newcommand{\nbox}{{\,\lower0.9pt\vbox{\hrule \hbox{\vrule height 0.2 cm \hskip 0.19 cm \vrule height 0.2 cm}\hrule}\,}}

\def\href#1#2{#2}

\begin{document}
\begin{titlepage}
\hfill CALT-TH 2016-009, IPMU16-0056\\
\vbox{
    \halign{#\hfil         \cr
           } % end of \halign
      }  % end of \vbox
\vspace*{15mm}
\begin{center}
{\Large \bf Gravitational Positive Energy Theorems \\ from Information Inequalities}

\vspace*{15mm}
\vspace*{1mm}
Nima Lashkari$^a$, Jennifer Lin$^{b}$, Hirosi Ooguri$^{c,d,e}$, \\ Bogdan Stoica$^c$, and Mark Van Raamsdonk$^f$
\vspace*{1cm}
\let\thefootnote\relax\footnote{$\mathrm{lashkari@mit.edu},\ \mathrm{jenlin@ias.edu},\ \mathrm{ooguri@theory.caltech.edu},\ \mathrm{bstoica@theory.caltech.edu},\ \mathrm{mav@phas.ubc.ca}$}

{${}^a$ Center for Theoretical Physics, Massachusetts Institute of Technology\\
77 Massachusetts Avenue, Cambridge, MA 02139, USA\\
\vspace*{0.1cm}
${}^b$ School of Natural Sciences, Institute for Advanced Study,
Princeton, NJ 08540, USA\\
\vspace*{0.1cm}
${}^c$ Walter Burke Institute for Theoretical Physics, \\ California Institute of Technology,
 Pasadena, CA 91125, USA\\
\vspace*{0.1cm}
${}^d$ Center for Mathematical Sciences and Applications and \\
Center for the Fundamental Laws of Nature, \\
Harvard University, Cambridge, MA 02138, USA\\
\vspace*{0.1cm}
${}^e$ Kavli Institute for the Physics and Mathematics of the Universe,
\\ University of Tokyo, Kashiwa 277-8583, Japan\\
\vspace*{0.1cm}
${}^f$ Department of Physics and Astronomy,
University of British Columbia, \\
6224 Agricultural Road,
Vancouver, B.C., V6T 1W9, Canada}

\vspace*{0.7cm}
%%\maketitle
\end{center}
\begin{abstract}
In this paper we argue that classical, asymptotically AdS spacetimes that arise as states in consistent ultraviolet completions of Einstein gravity coupled to matter must satisfy an infinite family of positive energy conditions. To each ball-shaped spatial region $B$ of the boundary spacetime, we can associate a bulk spatial region $\Sigma_B$ between $B$ and the bulk extremal surface $\tilde{B}$ with the same boundary as $B$. We show that there exists a natural notion of a gravitational energy for every such region that is non-negative, and non-increasing as one makes the region smaller. The results follow from identifying this gravitational energy with a quantum relative entropy in the associated dual CFT state. The positivity and monotonicity properties of the gravitational energy are implied by the positivity and monotonicity of relative entropy, which holds universally in all quantum systems.

\end{abstract}

\end{titlepage}

\vskip 1cm

\section{Introduction}

Consider a classical asymptotically AdS spacetime $M$
of $(d+1)$ dimensions associated with a state in some UV-complete theory of quantum gravity for which the low-energy effective description is Einstein gravity coupled to matter. According to the AdS/CFT correspondence, there is a corresponding state $|\Psi \rangle$ in a dual conformal field theory living on the
$d$-dimensional boundary spacetime $\partial {\cal M}$. For a spatial region $B$ of $\partial {\cal M}$, the Ryu-Takayanagi formula \cite{ryu2006holographic} (and its covariant generalization \cite{hubeny2007covariant}) relate the entanglement entropy of the CFT subsystem $B$ to the area of the minimal-area extremal surface $\tilde{B}$ in $M$ with boundary $\partial B$:
\be
\label{RT}
S_B(|\Psi\rangle) - S_B(|{\rm vac} \rangle) = {\rm Area}_M (\tilde{B}) -
{\rm Area}_{{\rm AdS}} (\tilde{B}) \, .
\ee
This connects a fundamental quantity in the quantum information theory of the CFT to a fundamental geometrical quantity in the dual gravitational theory.

In this paper, we make use of this result to derive another fundamental connection between quantum information theory and geometry. In this case, the information theoretic quantity is {\it quantum relative entropy}, a measure of distinguishability between a general state $\rho$ and some reference state $\sigma$. In our case, the state $\rho$ is the reduced density matrix $\rho^{\Psi}_B$ in our state $\Psi$ (generically time-dependent) for a ball-shaped subsystem $B$ of the CFT, and the reference state is the reduced density matrix $\sigma = \rho_B^{vac}$ for the same subsystem in the CFT vacuum state. We find that the relative entropy $S(\rho^\Psi_B||\rho_B^{vac})$ (reviewed in section 2.1 below) is related to a novel measure of energy associated with the spatial region
 $\Sigma_B$ between the boundary domain $B$ and the extremal surface
 $\tilde{B}$:
\be
\label{RelEnt}
S(\rho^\Psi_B||\rho_B^{{\rm vac}}) = {\rm Energy}_M (\Sigma_B)-
{\rm Energy}_{{\rm AdS}} (\Sigma_B) \, .
\ee
In the limit of small perturbations to AdS the region $\Sigma_B$ can be thought of as a Rindler patch of AdS, and the energy is the associated Rindler energy. The energy on the right hand side is covariantly defined (in section 2.3 below), and includes both matter and gravitational contributions. It can also be expressed as a purely geometrical quantity in terms of spacetime curvatures, so (\ref{RelEnt}) represents another element in the dictionary between quantum information and geometry.

A crucial property of relative entropy in quantum systems is that it is {\it positive} and {\it monotonic} (i.e. it increases if we consider a larger subsystem containing the original subsystem). Thus, our result (\ref{RelEnt}) gives rise to a new gravitational positive energy theorem: for any spacetime ${\cal M}$ described by a consistent theory of Einstein gravity coupled to matter, the background-subtracted energy on the right side of (\ref{RelEnt}) must be positive for all boundary subsystems $B$ and must increase if we move to a larger subsystem $B' \supset B$. Any spacetime $M$ which fails to satisfy this property is unphysical. Furthermore, any low-energy effective theory whose solutions violate the positivity and/or monotonicity properties cannot have a consistent UV completion: it lives in the swampland. Thus, the positivity and monotonicity of relative entropy in conformal field theories gives rise to novel constraints on physical asymptotically AdS spacetimes and on low-energy effective field theories.

\subsubsection*{Connection with Previous Work}

The results in this paper generalize a series of previous works investigating the gravitational interpretation of CFT relative entropy and the implications of its positivity and monotonicity. Relative entropy for holographic CFTs was originally introduced in \cite{blanco2013relative}, where the authors provided a direct holographic interpretation as a difference of bulk integrals on $\tilde{B}$ and $B$.

Constraints on the dual spacetimes from relative entropy positivity were considered at leading order in perturbations to pure AdS in \cite{Lashkari:2013koa,faulkner2014gravitation,swingle2014universality} and shown to be equivalent to Einstein's Equations linearized about the AdS background. The works \cite{Banerjee:2014oaa,Banerjee:2014ozp,Lin:2014hva,Lashkari:2014kda} discussed constraints beyond linear order. In particular the papers \cite{Lin:2014hva,Lashkari:2014kda} identified connections between relative entropy and bulk energy, and between relative entropy constraints and certain bulk energy conditions. In \cite{Lashkari:2015hha} this connection between relative entropy and bulk energy was established in general at second order in perturbations to pure AdS. The relative entropy at second order, known as Quantum Fisher Information, maps to a quantity known as the Canonical Energy associated with the spacetime region $\Sigma_B$. The papers \cite{faulkner2014gravitation,Lashkari:2015hha} relied upon a set of elegant results in classical gravitational theories due to Wald and various collaborators. This same technology is employed in the present paper to derive the result (\ref{RelEnt}) from the expression of \cite{blanco2013relative} involving boundary integrals.

Recently, it was pointed out in \cite{Jafferis:2015del}
that the relative entropy of nearby states in the CFT, in the sense that their gravity duals are different quantum states on the same background geometry,
is given by the relative entropy in the bulk. This result was used to prove the entanglement wedge reconstruction theorem
in \cite{Dong:2016eik}. These results show that the positivity and monotonicity of
the holographic relative entropy is automatically satisfied by states nearby the AdS
vacuum, nearby in the sense that these states consist of a few particle excitations on
the AdS vacuum without their backreaction to the geometry.  In this paper, we will
explore implications of the positivity and monotonicity of the relative entropy for
states whose bulk geometries are different from the AdS vacuum. Hence our results are orthogonal to those of \cite{Jafferis:2015del, Dong:2016eik}. We will show that
these information inequalities impose constraints on the bulk geometry, leading to
a certain set of positive energy conditions.

\subsubsection*{Outline}

In the next section of the paper we review the definition and properties of relative entropy in conformal field theories, recall some relevant background about energy in gravitational theories, and then make use of (\ref{RT}) to derive (\ref{RelEnt}), providing an explicit definition for the gravitational energy appearing there. We also provide an alternative derivation of (\ref{RelEnt}) in the case of time-symmetric geometries without using (\ref{RT}), employing a direct path-integral argument similar to the derivation of (\ref{RT}) in \cite{Lewkowycz:2013nqa}. In section 3, we discuss the implications of our result, describing the gravitational energy theorems that follow from positivity and monotonicity of relative entropy, and using these to derive some explicit geometrical constraints on consistent spacetimes. In section 4, we generalize a result from \cite{Lin:2014hva} showing that a certain differential operator acting on relative entropy (employing derivatives with respect to the ball radius $R$) can be identified with bulk matter energy density integrated over the extremal surface $\tilde{B}$ in the case of infinitesimal balls $B$. We find that the same differential operator applied to relative entropy for general balls $B$ is also dual to the integral of a certain bulk quantity over $\tilde{B}$ and derive an explicit expression for this. We conclude in section 5 with some further discussion and future directions.

\section{Relative Entropy}

In this section, we present a holographic description of the
relative entropy. After reviewing
the definition of the relative entropy in conformal field theory,
we will formulate the holographic dual of the relative entropy
in terms of the quasi-local energy associated to the region
between the boundary domain $B$ and the extremal surface
 $\widetilde{B}$ (Ryu-Takayanagi surface or its covariant generalization). We will also give a path integral derivation of
this holographic dual description along the lines
of the proof of the Ryu-Takayanagi formula by
 Lewkowycz and Maldacena.

\subsection{Relative Entropy in Conformal Field Theory}

For a general quantum system, relative entropy is a measure of distinguishability between a state $\rho$ and a reference state $\sigma$.\footnote{Orthogonal quantum states can always be perfectly distinguished using projective measurements. In other to account for this we define the relative entropy of two density matrices with $\sup\sigma\cap\ker\rho\neq 0$ to be infinite. Here $\sup\rho$ is the support of $\rho$ in the Hilbert space and $\ker\rho$ is its complement. A particular instance of infinite relative entropy is when $\sigma$ is pure with $\rho \ne \sigma$.} It is defined as\footnote{When $\sigma$ and $\rho$ commute they can be simultaneously diagonalized and quantum relative entropy becomes the Kullback-Leibler divergence of their eigenvalue vectors.}
\[
S(\rho||\sigma) = \tr(\rho \log \rho) - \tr(\rho \log \sigma) \, .
\]
If we add and subtract $\tr(\sigma\log\sigma)$ to the definition above one can recast relative entropy as a change in {\it free energy}  \cite{blanco2013relative}
\bea
\label{REdiff}
S(\rho\|\sigma)&=&\Delta \la H_\sigma\ra-\Delta S\\
&=&F_\sigma(\rho)-F_\sigma(\sigma)\nn\\
F_\sigma(\rho)&=&tr(\rho H_\sigma)-S(\rho)
\eea
where $H_\sigma$ is the ``modular Hamiltonian" of the reference state defined by $H_\sigma=-\log\sigma$ and $S(\rho) = - \tr(\rho \log \rho)$ is the Von Neumann entropy of $\rho$. In fact, quantum relative entropy is naturally interpreted as the extractable free energy of $\rho$ in a thermodynamic theory where $\sigma$ is the equilibrium state with respect to $H_\sigma$;
see appendix \ref{freeEnergy}.
Free energy is minimized on the equilibrium state; this implies that relative entropy is non-negative,
\be
\label{positivity}
S(\rho ||\sigma) \geq 0\,.
\ee
It vanishes if and only if $\rho$ is the same as the equilibrium state $\sigma$ .

It is often useful to consider the relative entropy $S(\rho_A||\sigma_A)$ for a subsystem $A$, where $\rho_A$ and $\sigma_A$ are the reduced density matrices for this subsystem. If $B$ is any larger subsystem $B \supset A$, we have
\be
\label{monotonicity}
S(\rho_A||\sigma_A) \le S(\rho_B||\sigma_B) \, ,
\ee
known as the monotonicity of relative entropy.

In this paper we consider the relative entropies when the reference state is the CFT vacuum and the regions are ball shaped.
%\footnote{The relative entropy for ball-shaped subsystems of a CFT on $d$-dimensional Minkowski space for two arbitrary states can be found by analytically continuing certain $2n$-point functions in $n$ \cite{Lashkari:2015dia}.}
In this case, the modular Hamiltonian appearing in (\ref{REdiff}) takes a simple form \cite{casini2011towards}. For a ball of radius $B$ centered at $x_0$ in the spatial slice perpendicular to the unit timelike vector $u^\mu$, the modular Hamiltonian is
\be
\label{defE}
H_B = \int_B \zeta_B^\mu T_{\mu \nu} \epsilon^\nu
\ee
where $\epsilon_\nu = \epsilon_{\nu \mu_1 \cdots \mu_{d-1}} dx^{\mu_1} \wedge \dots \wedge dx^{\mu^{d-1}} / (d-1)!$ is a volume form and $\zeta_B$ is the conformal Killing vector
\be
\label{defzeta}
\zeta_B^\mu = { \pi \over R} \left\{ [R^2 + (x-x_0)^2 + 2(u_\nu (x-x_0)^\nu)^2] u^\mu + [2 u_\nu (x - x_0)^\nu ](x - x_0)^\mu \right\} \, .
\ee
Thus, for ball-shaped regions in a general CFT state, the relative entropy to the vacuum is
\bea
\label{defRE}
S(\rho_B||\sigma_B) &=& \Delta \langle H_B \rangle - \Delta S_B \, ,
\eea
with $H_B$ given in (\ref{defE}). This is the object that we will translate directly to a bulk geometrical quantity in the case of a holographic CFT.

\subsection{Quasi-Local Energy}

In the next subsection, we will argue that the quantum relative entropy for a ball-shaped region in the CFT is related to the energy of a subsystem in the dual gravitational theory. First, it will be helpful to review some relevant background about energy in gravitational theories, following \cite{Iyer:1994ys,Wald:1999wa}.

It is believed that there are no local observables in a gravitational system.
However, if we can define a subspace $\Sigma$ of a Cauchy surface in a diffeomorphic invariant
way, we can formulate a notion of a quasi-local energy for $\Sigma$.
In the next subsection, we will consider $\Sigma$ defined as the part of a
Cauchy surface between
a boundary domain $B$ and the corresponding extremal
surface $\widetilde{B}$ (the Ryu-Takayanagi surface or its covariant generalization).

Consider a metric and a set of matter fields on the
$d$-dimensional surface $\Sigma$ described by a
Lagrangian density $L$, expressed as a $(d+1)$-form. To simplify notations, we will denote all
the fields by $g(x)$ (representing matter fields as well as
the metric).
By the variational principle,
\be
\label{eom}
\delta
L(g) = d \theta(\delta g) + {\rm equations \ of \ motion},
\ee
where $d$ acting on
$\theta(\delta g)$ on the right-hand side is the exterior derivative,
and $\theta$ is an $d$-form on $\Sigma$ that is linear in $\delta g$. We can think of $\theta(\delta g)$
as a one-form in the space of field configurations on $\Sigma$ and
define an associated symplectic form by
\be
\label{symplectic}
W(\delta_1 g, \delta_2 g) = \int_\Sigma \omega(\delta_1 g, \delta_2 g)
= \int_{\Sigma} \Big[ \delta_1 \theta(\delta_2 g) -
\delta_2 \theta(\delta_1 g) \Big].
\ee

Consider a vector field $\xi$ on $\Sigma$. It generates
an infinitesimal diffeomorphism on $\Sigma$. With an appropriate
boundary condition on $\partial \Sigma$, which we will specify below, the diffeomorphism is a symmetry of the subsystem on
$\Sigma$, in which case we can define a Hamiltonian $H_\xi$,
which generates the diffeomorphism as a symplectic
transformation on $g(x)$ as
\be
\label{varH}
  \delta H_\xi = \int_\Sigma \omega(\delta g, {\cal L}_\xi g) \;.
\ee
Here, ${\cal L}_\xi g$ is the Lie derivative of $g$ with respect to
the vector field $\xi$.
By definition (\ref{symplectic}),
$\omega(\delta g, {\cal L}_\xi g) = \delta \theta({\cal L}_\xi g)
- {\cal L}_\xi \theta (\delta g)$.
Since ${\cal L}_\xi \theta = \xi \cdot d \theta + d (\xi \cdot \theta)$ and $d \theta = \delta L$ by the equations of motion,
\bea
\label{infinitesimal_Hamiltonian}
 \delta H_\xi & = & \int_\Sigma
 \Big[ \delta\theta ({\cal L}_\xi g) - \xi \cdot \delta L
 - d(\xi \cdot \theta(\delta g) ) \Big] \nonumber \\
 & = & \int_\Sigma \delta J_\xi
 - \int_{\partial \Sigma} \xi \cdot \theta(\delta g),
 \eea
 where
\be
\label{JefD}
J_\xi = \theta ({\cal L}_\xi g) - \xi \cdot L.
\ee
This is the Noether current form associated with the diffeomorphism.

If we can  find a $d$-form $K(g)$ on the boundary $\partial \Sigma$ such that,
 \be
 \label{integrate_theta} \delta (\xi \cdot K) = \xi \cdot \theta(\delta g) ~~~
 {\rm on \ \partial \Sigma}\,,
 \ee
 we can integrate (\ref{infinitesimal_Hamiltonian}) in the field
 configuration space to define,
 \be
 \label{Hamiltonian}
  H_\xi = \int_\Sigma J_\xi - \int_{\partial \Sigma} \xi \cdot K.
 \ee
 Since $\omega = \delta \theta$, the boundary term $K$ can be found if $\xi$ and $\omega$ satisfy the integrability condition,
\be
\label{integrability}
 \int_{\partial\Sigma} \xi \cdot \omega(\delta_1 g, \delta_2 g) = 0,
 \ee
 for any infinitesimal variations $\delta_1 g$ and $\delta_2 g$
 allowed on $\partial \Sigma$.
 In this case, $H_\xi$ gives a natural definition of a quasi-local
 energy for the region $\Sigma$ with respect to the vector field
 $\xi$. (It is useful to remember that, in a simple system
 $L = k(dq/dt) - V(q)$, the
 Hamiltonian $H$ for $t$-translation
 is defined as $H = p \ dq/dt - L$, where $p = dk/d(dq/dt)$. The Hamiltonian $H_\xi$ defined
 here is its natural generalization.)

 If $\Sigma$ is the entire Cauchy surface that asymptotes to
 the AdS boundary and if $\xi$ approaches one of the
  conformal Killing  vectors on the boundary,
  $H_\xi$ is the holographic dual to the generator of the
  conformal transformation on the boundary CFT.
  In this case, the boundary term $K$ is the standard
  Gibbons-Hawking term for the pure gravity and its
  appropriate generalization for a general gravitational
  system, which one can identify
  using the holographic renormalization
  group formalism.

The conservation of the current $J_\xi$ can be easily checked as,
\be
 d J_\xi = d \theta({\cal L}_\xi g) - d(\xi \cdot L)
 = {\cal L}_\xi \cdot L - d(\xi \cdot L) = 0,
 \ee
 where we used $\delta L = d\theta(\delta g)$
by the equations of motion. 
%If the region $\Sigma$ is
%contractible, which is the case considered in this paper,

Furthermore \cite{Iyer:1994ys}, we can find a $(d-1)$-form $Q_\xi$ such that on shell,
\be
\label{JdQ}
  J_\xi = d Q_\xi.
\ee
Thus, the Hamiltonian $H_\xi$ can be expressed as
the integral over the boundary,
\be
\label{boundary_hamiltonian}
H_\xi = \int_{\partial \Sigma} \Big[ Q_\xi - \xi \cdot K \Big].
\ee
This means that $H_\xi$ depends on the vector field $\xi$
only through its properties near the boundary $\partial \Sigma$. For the case of Einstein gravity with cosmological constant, explicit expressions for all the quantities appearing in this section are given in appendix \ref{Waldappendix}.

\subsection{Holographic Relative Entropy}

We will now see that the gravitational quantity associated with the CFT relative entropy for a ball-shaped region coincides with a particular gravitational Hamiltonian as defined in the previous section.

Consider a gravitational solution in the bulk that is dual to
a state in the CFT. For a domain $B$ on the boundary, let $\Sigma$ be a spacelike surface between
 $B$ and the corresponding bulk extremal
surface $\widetilde{B}$.
We will show that there is a choice of a vector field
$\xi$ in a neighborhood of $\Sigma$ such that the difference of the
quasi-local energy $H_\xi$ for the gravitational solution
minus the energy for the vacuum AdS geometry gives the relative entropy
between the state $\rho_B$ dual to our gravitational solution
and the state $\rho_B^{\rm vac}$ for the vacuum.

In some sense, we already have a holographic description for the relative
entropy $S(\rho_B ||\rho_B^{\rm vac})$ since it is equal to
$\Delta \langle H_B \rangle - \Delta S$ as explained in section 2.1 and since
both the expectation value of the modular Hamiltonian  $\langle H_B \rangle$
and the entanglement entropy $S$ have holographic counterparts.
To the leading order
 in large $N$,
the covariant holographic entanglement entropy formula shows that
\be
 \Delta S = \frac{1}{4G_N}\Delta {\rm Area}(\tilde{B}).
 \ee
As explained in \cite{faulkner2014gravitation}, this formula implies directly that the CFT stress tensor expectation value is related to the asymptotic metric via the usual relation\footnote{The derivation proceeds by applying the entanglement first law to an infinitesimal ball. In this case, the variation in the entanglement entropy is related to the asymptotic metric (since it is proportional to the area of an extremal surface near the AdS boundary), while the modular Hamiltonian expectation value is related to stress tensor expectation value at a point.}
\be
\label{holoT}
\Delta \langle T_{\mu \nu} \rangle = \Delta T_{\mu\nu}^{grav} \equiv {d \ell^{d-3} \over 16 \pi G_N}
\, \Gamma_{\mu \nu}(x,z=0)\ \ ,
\ee
where $\Gamma_{\mu \nu}$ is defined by the Fefferman-Graham description of the metric for M,
\be
\label{FGmetric}
ds^2 = {\ell^2 \over z^2}\left(dz^2 + dx_\mu dx^\mu + z^{d-1} \Gamma_{\mu \nu}(z,x) \right) \, .
\ee
Therefore, the relative entropy can be expressed as an integral over $B$ and
$\widetilde{B}$ as,
\be
\label{holographic_one}
 S(\rho_B ||\rho_B^{\rm vac})
 ={d \ell^{d-3} \over 16 \pi G_N} \int_B \zeta_B^\mu
\, \Gamma_{\mu \nu}(x,z=0) \epsilon^\mu
- \frac{1}{4G_N}\ \Delta {\rm Area}(\widetilde{B}).
\ee

What we would like to do is to relate (\ref{holographic_one}) to
the quasi-local energy $H_\xi$ defined in the previous subsection
for some choice of $\xi$. In this way, we can translate the positivity
and monotonicity of the relative entropy to conditions on the quasi-local energy.

The choice of $\xi$ may be motivated by the result \cite{Lashkari:2015hha} that to quadratic order in perturbation theory, the relative entropy maps to the bulk energy associated with a Killing vector (here given in Fefferman-Graham coordinates)
\bea
\label{defxi}
\xi_B &=& { \pi \over R}\left\{ [R^2 - z^2 + (x-x_0)^2 + 2(u_\nu (x-x_0)^\nu)^2] u^\mu + [2 u_\nu (x - x_0)^\nu ](x - x_0)^\mu \right\} \partial_\mu \cr
 && \qquad \qquad + { \pi \over R} \left\{ u_\nu (x-x_0)^\nu z \right\} \partial_z     \, ,
\eea
where $u$ is the timelike orthogonal vector to the spatial slice in which the ball resides. The vector $\xi_B$ reduces to the conformal Killing vector $\zeta_B$ at the boundary and vanishes on the extremal surface $\tilde{B}$.

For general asymptotically AdS spacetimes, there are no Killing vectors, but we can find a vector $\xi$ that behaves in the same way near $B$ and $\widetilde{B}$ as the Killing vector behaves near these surfaces in pure AdS. Specifically, we require\footnote{The second condition is that the vector satisfies the Killing equation ${\cal L}_\xi g = 0$ up to order $z^{d-3}$. Alternatively, we can require that in Fefferman-Graham coordinates, $\xi$ agrees with (\ref{defxi}) up to corrections of order $z^{d+1}$.}
\bea
\xi^a|_{B} &=& \zeta^a_B, \label{x3} \\
\nabla_{(a}\xi_{b)}|_{z \to 0} &=& {\cal O}(z^{d-2}), \label{x4} \\
\nabla^{[a}\xi^{b]}|_{\tilde B} &=& 2\pi n^{ab}, \label{x1} \\
\xi|_{\tilde B} &=& 0, \label{x2}
\eea
where $n^{ab} = n_1^a n_2^b - n_2^a n_1^b$ is the binormal unit vector to $\widetilde B$ and $\zeta_B$ is the conformal Killing vector \eqref{defzeta}. As we show in appendix \ref{a2}, it is always possible to find such $\xi$. The choice of $\xi$ is not unique since it is unconstrained away from $B$ and $\tilde{B}$, but the value of $H_\xi$ will not depend on the detailed behavior of $\xi$ in the interior of $\Sigma$ since $H_\xi$ can be expressed as a boundary integral as in (\ref{boundary_hamiltonian}). We will give one explicit construction for $\xi$ in section 4.

If $\xi$ satisfies these boundary conditions (\ref{x3}) - (\ref{x2}),
we can show that $K$, as defined in the previous section, exists, and that
 \bea
 \label{holographic_energy}
   \Delta  \int_{B} \Big[ Q_\xi - \xi \cdot K \Big]
   &=& {d \ell^{d-3} \over 16 \pi G_N} \int_B \zeta_B^\mu
\, \Gamma_{\mu \nu}(x,z=0) \epsilon^\mu ,  \\
\label{holographic_entropy}
\Delta  \int_{\widetilde{B}} \Big[ Q_\xi - \xi \cdot K \Big]
&=&
 \frac{1}{4G_N}\  \Delta {\rm Area}(\widetilde{B}).
 \eea
These results allow us to rewrite (\ref{holographic_one}) as a difference of the quasi-local energy,
\be
\label{Hdiff}
  S(\rho_B || \rho_B^{({\rm vac})})= H_\xi(M) - H_\xi({\rm AdS}),
\ee
where $H_\xi$ is the Hamiltonian (\ref{boundary_hamiltonian}) associated with the vector field $\xi$. Thus we can identify $H_\xi$ as the ``novel measure of energy'' discussed in the introduction,
\be
   {\rm Energy}_M (\Sigma) = H_\xi(M),
   \ee
for the region $\Sigma$ of the Cauchy surface of the spacetime $M$.

To show eq. \eqref{holographic_entropy},
note that $\xi \cdot K$ vanishes on the surface $\tilde B$ because $\xi$ vanishes there by the boundary condition \eqref{x2}\,. Further, for the theories we are considering (with Einstein gravity coupled to matter, where the matter couplings do not involve curvatures), $Q_\xi$ may be chosen to take the form \cite{Iyer:1994ys}\footnote{From \cite{Iyer:1994ys}, the most general form of $Q$ in this case is $Q_\xi = -{1 \over 16 \pi G_N} \nabla^a \xi^b \epsilon_{ab} + W^a \xi_a + Y(\phi, {\cal L}_\xi \phi) + dZ$; however, the $Z$ terms is a total derivative which does not affect the integral of $Q$ on a boundary, $Y$ can be removed by making use of the ambiguity $\theta \to \theta + d Y$ in (\ref{eom}), and $W$ can be removed by the freedom $Q \to Q + \xi \cdot \mu$ and $\theta \to \theta + \delta \mu$ which corresponds to adding a total derivative $d \mu$ to the Lagrangian form. We will assume that these choices have been made to remove the possible extra terms.}
\be
\label{defQ}
Q_\xi = -{1 \over 16 \pi G_N} \nabla^a \xi^b \epsilon_{ab} \;.
\ee
Here $\epsilon_{ab}$, defined in appendix \ref{Waldappendix}, is defined such that its contraction with orthogonal unit vectors $n_1$ and $n_2$ gives the volume form in the perpendicular subspace. The boundary condition \eqref{x1} for $\xi$ then implies that $Q_\xi$ evaluated on $\tilde{B}$ is $1/(4 G_N)$ times the volume form on $\tilde{B}$, so we have
\be \label{eq23}
\int_{\widetilde{B}}  Q_\xi = -{1 \over 16 \pi G_N} \int_{\tilde B} \nabla^a \xi^b \epsilon_{ab} = \frac 1{4G_N} {\rm Area}(\tilde{B})\,,
\ee
as desired.

To show eq. \eqref{holographic_energy}, consider the infinitesimal version of the left side,
\[
\int_B (\delta Q_\xi - \xi \cdot \theta(g,\delta g)) \; .
\]
In this expression, the terms that survive the limit when the cutoff surface $B$ approaches the boundary involve only the leading deviations from the pure AdS metric in the asymptotically AdS geometry $M$. Furthermore, the expression is linear in these perturbations, which can be represented explicitly by the tensor $\Gamma_{\mu \nu}(x,z=0)$ appearing in (\ref{FGmetric}). In \cite{Faulkner:2013ica}, it was shown explicitly that for a Fefferman-Graham description of the metric, these linear perturbations satisfy\footnote{In that calculation, the expression for $\xi$ in Fefferman-Graham coordinates was assumed to be that of the Killing vector in pure AdS. The condition  (\ref{x4}) ensures that for the more general $\xi$ vectors we are considering, no additional terms appear in the expression below.}
\be\label{faulkner}
\int_B (\delta Q_\xi - \xi \cdot \theta(g,\delta g)) = {d \ell^{d-3} \over 16 \pi G_N} \int_B \zeta_B^\mu
\, \delta \Gamma_{\mu \nu}(x,z=0) \epsilon^\mu.
\ee
To recover \eqref{holographic_energy}, we can simply integrate this expression on a one-parameter family of metrics from pure AdS to the desired spacetime. Since the first term on the left and the term on the right give results that are independent of which path through the space of metrics we choose, this must also be true for the term involving $\theta$. This establishes the existence of $K$ as in (\ref{integrate_theta}),\footnote{Alternatively, the existence of $K$ follows from the integrability condition (\ref{integrability}), which was argued in \cite{Hollands:2005wt} based on the vanishing of $\omega$ at the AdS boundary.} and the result is precisely \eqref{holographic_energy}.

In calculating the difference in \eqref{holographic_energy}, we require a regularization procedure in which quantities are calculated on a regularization surface away from the boundary, the results for the two spacetimes are subtracted, and then the surface is taken to the boundary. It is useful to note that the result does not depend on the precise way in which these surfaces are chosen. This follows because the infinitesimal variation appearing on the left side in (\ref{faulkner}) satisfies
\be
d(\delta Q_\xi - \xi \cdot \theta(g,\delta g)) = 0
\ee
on shell to leading order in perturbations to AdS. Thus, by Stokes' theorem, for two choices of surface $B$ and $B'$, we have
\be
\int_B (\delta Q_\xi - \xi \cdot \theta(g,\delta g)) - \int_{B'} (\delta Q_\xi - \xi \cdot \theta(g,\delta g)) = {\cal O}(\delta g^2).
\ee
The non-linear perturbations on the right appear only at higher orders in the Fefferman-Graham expansion and do not contribute in the limit where the surface is taken to the boundary.

A particularly convenient choice of surface is the $z=\epsilon$ surface in Fefferman-Graham coordinates. For this choice, direct calculation shows that the first term on the left side in (\ref{faulkner}) equals the right side, the $\theta$ term doesn't contribute, and we have
\be
 S(\rho_B || \rho_B^{({\rm vac})}) = \Delta ( \int_{B_{FG}} Q_\xi - \int_{\tilde{B}} Q_\xi ) = \Delta \int_{\Sigma_{FG}} J_\xi \,,
\ee
where the subscript $FG$ indicates that the $z=\epsilon$ surface in Fefferman-Graham coordinates is to be used when performing the subtraction. The simple result $H_\xi = \int_{\Sigma_{FG}} J_\xi$ shows that the Hamiltonian $H_\xi$ also has the conventional interpretation as the conserved charge associated with the diffeomorphism symmetry generated by $\xi$.

\subsection{Path Integral Derivation}

In this section, we derive the gravity dual of relative entropy for time-independent states without assuming the Ryu-Takayanagi formula. Instead, similar to the method presented in \cite{Lewkowycz:2013nqa} we assume AdS/CFT and bulk equations of motion.

It was shown in \cite{Lashkari:2014yva} that there exists a $Z_n$-symmetric replica trick that computes the relative entropy of excited states with respect to vacuum reduced to ball-shaped regions. In this replica trick, relative entropy is found from the analytic continuation of R\'enyi relative entropies:
\bea
S(\rho\|\sigma)=\lim_{n\to 1}S_n(\rho\|\sigma),
\eea
where
\bea\label{Sn}
&&S_n(\rho\|\sigma)=\frac{1}{n-1}\log \lb\frac{tr\lb\tilde{\rho}^n\rb}{tr(\rho)^ntr(\sigma)^{1-n}}\rb,\nn\\
&&\tilde{\rho}=\sigma^{\frac{1-n}{2n}}\rho\sigma^{\frac{1-n}{2n}}.
\eea
Assuming analyticity, the limit $n\to 1$ corresponds to taking a derivative with respect to $n$:
\bea\label{replica}
S(\rho\|\sigma)=\p_n\log tr(\tilde{\rho}^n)\Big|_{n=1}-\log tr(\rho)+\log tr(\sigma).
\eea
As we will see, $tr(\tilde{\rho}^n)$ in conformal field theory is a one-sheeted partition function. Therefore, from the operator-state correspondence we know that R\'enyi relative entropies are functions of Euclidean correlators.
% that has a gravity dual with metric $g(n)$ and matter fields $\phi(n)$.

\begin{figure}
\centering
\includegraphics[width=0.8\textwidth]{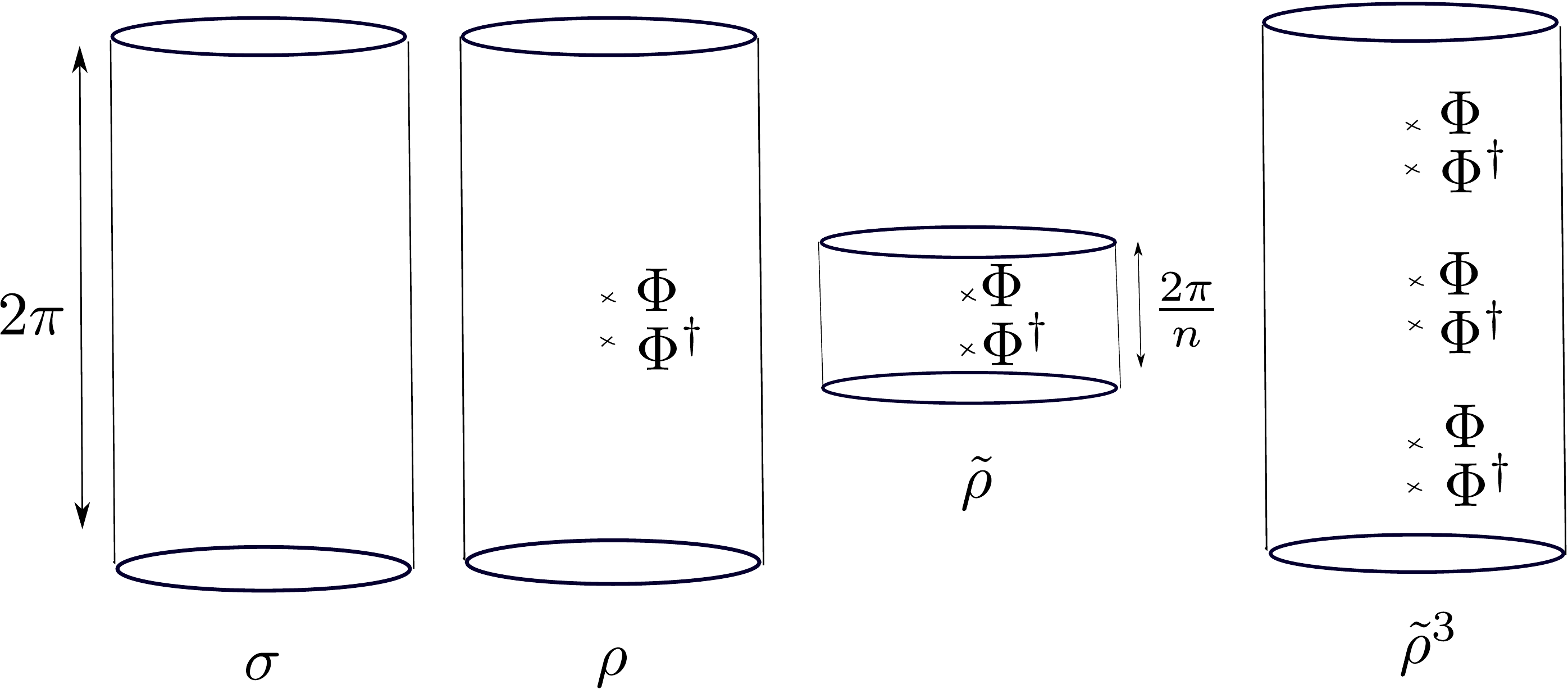}\\
\caption{\small{The two Euclidean path-integrals on the left prepare the density matrix of a spherical subsystem in a CFT in vacuum and an arbitrary state, respectively $\sigma$ and $\rho$. The path-integrals appearing in the definition of R\'enyi relative entropies are of the type on the right.}}
\label{fig1}
\end{figure}

Consider the vacuum state in a $d$-dimensional conformal field theory reduced to a ball of radius $R$. There exists a unitary transformation that maps this density matrix to a thermal state on hyperbolic space $H^{d-1}$: $\sigma\sim P e^{-2\pi H_B}$, where $H_B$ is the Hamiltonian on $H^{d-1}$ defined in (\ref{defE}). Up to normalization, this density matrix is prepared using a Euclidean path-integral on $H^{d-1}\times (0,2\pi)$.
%where $H_0$ is the Hamiltonian that generates time translation in the hyperbolic space {\bf where is $H_0$ here?}, and is the same as (\ref{defE}) on $t=0$ surface.
The operator-state correspondence in conformal field theory implies that an arbitrary excited state reduced to the same ball is $\rho\sim P e^{-\int_0^{2\pi}d\tau H(\tau)}$, where $H(\tau)$ is $H_B$ everywhere except at two points. At $\tau=(\pi\pm \ep)$ we need to insert in the path-integral the operators $\Phi$ and $\Phi^\dagger$ that create and annihilate the global state. Here $R/\ep$ is the infrared cut-off of the theory; see appendix \ref{conformaltrans}.
%Let's start with the reduced density matrix of a spherical subsystem in an excited state. A Euclidean path-integral with operator insertions at $\tau $ prepares this density matrix.
%%A conformal transformation maps this to the Rindler wedge with $x>0$.
%
%If the state is made from the action of a primary operator at $\tau\to -i\infty$ then it is obvious how it changes under a conformal transformation.
%% followed by a Weyl transformation.
%
%%\section{Information theorist's proof}
%%
Figure \ref{fig1} shows that the operator $\tilde{\rho}$ has an expression in terms a Euclidean path-integral on hyperbolic space with Euclidean time-direction $\tau$ in the interval $(\pi(1-1/n),\pi(1+1/n))$:
 \bea\label{Renyirel}
%&&S_n(\rho\|\sigma)=\frac{1}{n-1}\log \lb\frac{tr\lb\tilde{\rho}^n\rb}{tr(\rho)^ntr(\sigma)^{1-n}}\rb\nn\\
&&\tilde{\rho}=\sigma^{\frac{1-n}{2n}}\rho\sigma^{\frac{1-n}{2n}}\sim P e^{-\int_{\pi(1-1/n)}^{\pi(1+1/n)} d\tau H(\tau)}.\nn
\eea
Sewing $n$-copies of $\tilde{\rho}$ together we find
\bea
&&tr(\tilde{\rho}^n)=tr(\sigma) \la \prod_{i=1}^n \Phi \Phi^\dagger\ra_{H^{d-1}\times S^1},
\eea
where the periodicity of $S^1$ is $2\pi$.
%Note that as opposed to the entanglement entropy replica trick, the partition functions that correspond to R\'enyi relative entropies have only one-sheet for all $n$.

According to AdS/CFT, the traces of holographic CFT states on the gravity side are found by evaluating the gravitational on-shell action over the Euclidean geometry and matter fields  dual to the state: $tr(\rho)=e^{-I_E(g(\rho))}$.
The on-shell action has a bulk piece and a boundary piece defined in (\ref{integrate_theta}):
\bea
&&\log tr \rho=-\int_{\mathcal M}L_E-\int_{\p{\mathcal M}}K.\nn
\eea
For Dirichlet boundary conditions at infinity, $K$ is the familiar Gibbons-Hawking type term one adds in holographic renormalization to ensure that the equations of motion are satisfied in the bulk.

\begin{figure}
\centering
\includegraphics[width=0.8\textwidth]{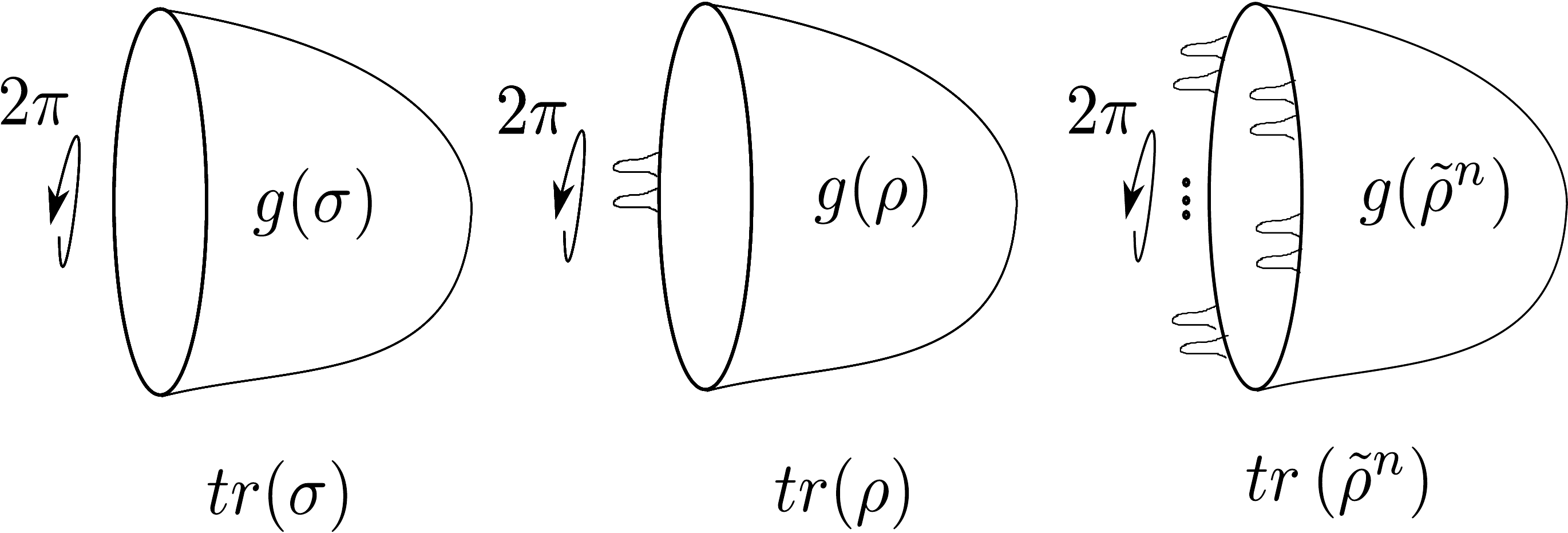}\\
\caption{\small{The bulk version of the replica trick in figure \ref{fig1}. Geometries on the left are dual to vacuum and excited state density matrices, respectively $\sigma$ and $\rho$ . The bulk configuration on the right prepares our quantity of interest in the definition of R\'enyi relative entropies.}}
\label{fig2}
\end{figure}

The CFT path-integrals on $H^{d-1}\times S^1$ can be extended into the bulk as illustrated in figure \ref{fig2}. The Euclidean metric dual to vacuum density matrix is the Euclidean hyperbolic black hole:
\bea
ds^2=\lb \frac{\rho^2}{R^2}-1\rb d\tau^2+\lb \frac{\rho^2}{R^2}-1\rb^{-1}d\rho^2+\rho^2 ds_{H^{d-1}}^2.
\eea
Using the proper distance from the horizon $r=\int \frac{R d\rho}{\sqrt{\rho^2-R^2}}$ as the radial coordinate, the metric takes the form
\bea
ds^2&=& \alpha(r)^2 d\tau^2+dr^2+R^2\cosh^2(r/R) ds_{H^{d-1}}^2,
\eea
where $\alpha(r)=R\sinh(r/R)=r+O(r^3)$ near the horizon at $r=0$. The Killing vector field $\p_\tau$ of the hyperbolic black hole geometry is the Euclidean analogue of $\xi_B$ in (\ref{defxi}).

The gravity dual to $tr(\tilde{\rho}^n)$ is the cigar geometry that is the solution to the bulk equations of motion with $Z_n$-symmetric boundary conditions $\tau\to\tau+2\pi/n$ on $H^{d-1}\times S^1$ at infinity. Following \cite{Lewkowycz:2013nqa}, we demand the solution $g(\tilde{\rho}^n)$ to remain $Z_n$ symmetric in the bulk.
The cigar caps off smoothly in the bulk where the $S^1$ circle shrinks to a point at a co-dimensional two surface we call $\tilde{B}(n)$. This surface is the fixed point of the action of $Z_n$ in the bulk. We can set up Gaussian normal coordinates near $\tilde{B}(n)$ analogous to the hyperbolic black hole,
\bea\label{GaussianNormal}
&&ds^2=\alpha^2d\tau^2+dr^2+2\beta_i d\tau dx^i+g_{ij} dx^i dx^j,\nn\\
%&&g_{ij}=h_{ij}+r \cos\tau K_{ij}^1+r\sin\tau K_{ij}^2+O(r^2)\nn\\
&&\alpha(r,\tau,n)=r+O(r^3),\qquad b_i(r,\tau,n)=O(r^2),
\eea
where $x^i$ are the directions along $\tilde{B}(n)$. In these coordinates $\tilde{B}(n)$ sits at $r=0$ where the vector field $\xi=\p_\tau$ vanishes.

\begin{figure}
\centering
\includegraphics[width=0.8\textwidth]{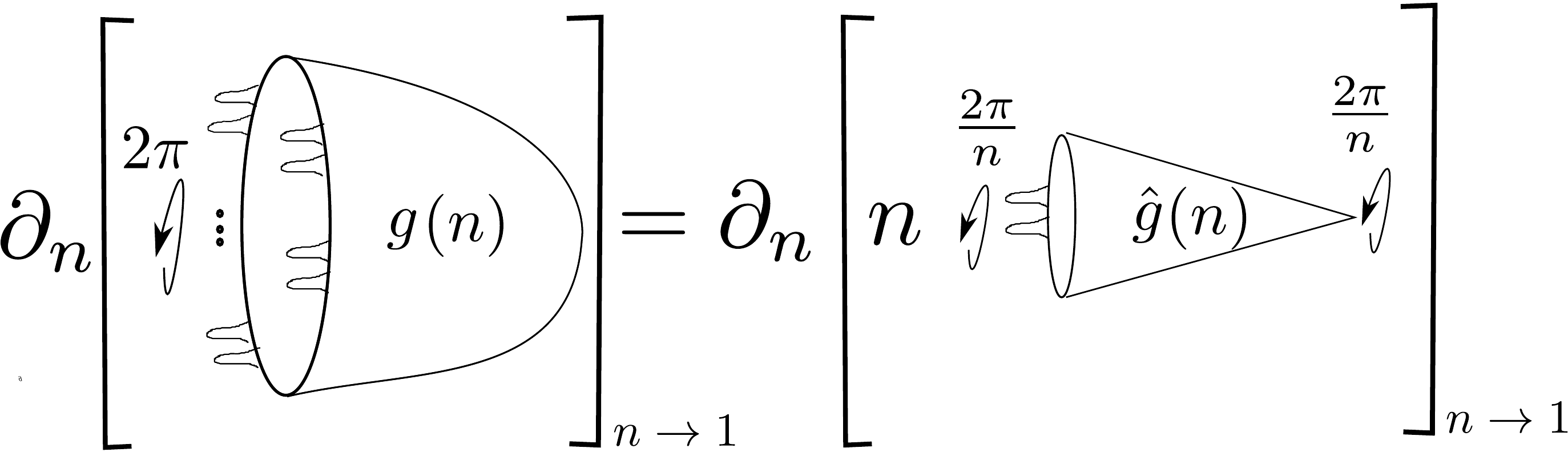}\\
\caption{\small{The analytic continuation of geometries to non-integer $n$ near one.}}
\label{fig3}
\end{figure}

We need to analytically continue $tr(\tilde{\rho}^n)$ in $n$. We define the analytic continuation to non-integer $n=1+\delta n$ to be
\bea\label{analytic}
\log tr(\tilde{\rho}^n)=-n I(\hat{g}(n))=-n\lb \int_{cone}L_E(\hat{g}(n))+\int_{\p (cone)}K(\hat{g}(n))\rb,
\eea
where $\hat{g}(n)$ is the solution to the bulk equations of motion on a cone with periodicity $2\pi/n$ with boundary conditions corresponding to $tr(\tilde{\rho})$ at infinity. The cone condition can be imposed by putting a brane at $r=0$ that creates an opening angle $2\pi/n$ around it.  The action in (\ref{analytic}) should include neither the brane action nor any contributions from the tip of the cone. The expression in (\ref{analytic}) can be alternatively interpreted as an off-shell smooth geometry with the same boundary conditions as $tr(\tilde{\rho}^n)$. The configuration $g(\tilde{\rho}^n)$ is on-shell which implies that its action differs from the proposed analytic continuation at order $(\delta n)^2$. This will not be an issue since relative entropy is derived from the coefficient of the term linear in $\delta n$; see figure \ref{fig3}. Now, we are ready to perform the analytic continuation in $n$:
\bea\label{relReplica}
S(g(\rho)\|g(\sigma))=-\p_n I(\hat{g}(n))\Big|_{n\to 1}+\log tr\sigma.
\eea

The definition of the vector field $\xi=\p_\tau$ near $\tilde{B}$ in (\ref{GaussianNormal}) can be extended everywhere in the bulk, leading to a foliation of the Euclidean geometry by surfaces of constant $\tau$. We demand $\xi$ to approach the generator of Euclidean time-translations on $H^{d-1}\times S^1$ at infinity, which is the Euclidean analogue of $\zeta_B$. Given any foliation of this type, we can compute the on-shell action of $\hat{g}(n)$ using the Hamiltonian that generates the flow along the vector field $\xi$  over the cone:
\bea
I(\hat{g}(n))=\int_{\pi(1-1/n)}^{\pi(1+1/n)}d\tau\: \lb \int_{\Sigma(\tau)}\:
\xi \cdot L(\hat{g}(n))+\int_{\p\Sigma(\tau)}\xi \cdot K(\hat{g}(n))\rb.
\eea
 Changing $n$ changes the periodicity both at $r=0$ and $r\to\infty$. Let us cut the cone open at $\tau=\pi(1-1/n)$ and represent the on-shell action with the short-form notation: $\int_{\pi(1-1/n)}^{\pi(1+1/n)}L$. Then,
\bea\label{pnI}
\p_nI(\hat{g}(n))\Big|_{n=1}&=&\lb\frac{d}{dn}\int_{\pi(1-1/n)}^{\pi(1+1/n)}L-\int_{\pi(1-1/n)}^{\pi(1+1/n)} \p_n L\rb_{n=1}\\
&+&\lb\frac{d}{dn}\int_{\pi(1-1/n)}^{\pi(1+1/n)}K-\int_{\pi(1-1/n)}^{\pi(1+1/n)}\p_n K\rb_{n=1}.\nn
\eea
%The contribution of the total derivative terms in the expression are given by:
%\bea
%\frac{d}{d\tau}\int_0^{\bar{\tau}}L=\int_{\Sigma(\bar{\tau})}\xi \cdot L.
%\eea
One can use the bulk equation of motion to write the terms on the right hand side in (\ref{pnI}) as boundary terms
\bea
\int_{\tau_1}^{\tau_2} \p_nL=\Theta(\p_ng\large|_{\tau_1})-\Theta(\p_ng\large|_{\tau_2})+\int_{\p{\mathcal M}}\Theta(\p_ng).
\eea
As a result
\bea
\p_nI(\hat{g}(n))\Big|_{n=1}=-2\pi\lb \int_{\Sigma(0)}\lb \xi \cdot L(g)-\Theta({\mathcal L}_\xi g)\rb+\int_{\p\Sigma(0)}\xi \cdot K(g)\rb,
\eea
 where we have used $\hat{g}(1)=g$, and the definition of $K(g)$ in (\ref{integrate_theta}).
 Note that the term $J_\xi=\Theta({\mathcal L}_\xi g)-\xi \cdot L(g)$ is the Hamiltonian that generates the flow along the $\xi$ vector field.
Therefore,
\bea
S(\rho\|\sigma)&=&-2\pi\lb\int_{\Sigma(0)}J_\xi+\int_{\p\Sigma(0)}\xi \cdot K \rb_{g(\rho)}+2\pi\lb\int_{\Sigma(0)}J_\xi+\int_{\p\Sigma(0)}\xi \cdot K \rb_{g(\sigma)},\nn
\eea
where we have used the fact that $\xi$ is a Killing vector in vacuum AdS.
%
%*****
%
%where $\Delta A\equiv A(\phi(\lambda))-A(\phi(0))$.
In order to compare with the Lorentzian result in the previous subsection one has to make the Wick rotation $\tau= i t$. In the Euclidean geometry on the $\tau=0$ surface this sends $\xi \rightarrow -i\xi$ and $L_E=-i L$. As before, we find that the relative entropy is the change in the phase space Hamiltonian associated with vector field $\xi$:
\bea
&&S(\rho\|\sigma)=H_\xi(g(\rho))-H_\xi(g(\sigma)),\nn\\
&&H_\xi(g)=2\pi\lb\int_{\Sigma(0)}J_\xi(g)+\int_{\p\Sigma(0)}\xi \cdot K(g) \rb.
\eea

%\subsection{Eternal black hole}
%Consider an entangled state of two CFTs on hyperbolic disks that is dual to a perturbation of the thermal field double (TFD) state.  The gravity dual of this state is a perturbation of the eternal hyperbolic black hole. Let us denote the subsystem corresponding to half of the left and right hyperbolic CFTs with $LR$. In appendix \ref{Eternal} we find the modular operator corresponding to subsystem $LR$ in the thermal field double state. We further extend the generator of this boundary modular flow to the bulk conformal Killing vector field $\xi$ in the eternal black hole geometry. From the arguments presented above,
%the gravity dual of relative entropy in subsystem $LR$ in a perturbed state with respect to the eternal black hole is
%\bea\label{halfhalf}
%&&S_{LR}(g\|g_{TFD})=H_\xi(g)-H_\xi(g_{TFD})\nn\\
%&&H_\xi=\int_\Sigma J_\xi-\int_{\p\Sigma}\xi\cdot K
%\eea
%where $\xi$ is chosen to satisfy the requirements (\ref{x3}) to (\ref{x2}). \footnote{It is worth noting that our result is valid only if the area of the extremal surface is the entanglement entropy of the state. Hence, at this point there is no clear connection between (\ref{halfhalf}) and the ``complexity" quantity introduced in \cite{susskind2016computational}.}

\section{Implications}

Using our identification of relative entropy with the vacuum-subtracted gravitational energy $\Delta H_\xi$, we now explore the implications of the relative entropy inequalities for spacetime geometry and gravitational physics.

\subsection{Positive energy theorems for gravitational subsystems}

We have seen that in any example of AdS/CFT for which the Ryu-Takayanagi formula (\ref{RT}) holds, the relative entropy for a ball-shaped region $B$ in the CFT is dual to the gravitational energy (\ref{Hamiltonian}) or (\ref{boundary_hamiltonian}) associated with $\xi_B$. When combined with relative entropy inequalities (\ref{positivity}) and (\ref{monotonicity}) that hold for all quantum systems, this result leads immediately to new positive energy theorems for asymptotically AdS spacetimes.

Specifically, the positivity of relative entropy (\ref{positivity}) implies that for any geometry $M$ associated with a consistent CFT state, the vacuum-subtracted energy $H_{\xi_B} - H_{\xi_B}^{AdS}$ associated with the subsystem $\Sigma_B$ between $B$ and $\tilde{B}$ must be positive for any ball-shaped boundary region in any Lorentz frame. The monotonicity of relative entropy implies further that for any two balls $B'$ and $B$, with $B$ in the domain of dependence of $B'$ the energy associated with $\Sigma_{B'}$ must be larger than the energy associated with $\Sigma_{B}$.

These results are much more detailed than the usual positive energy theorems  \cite{AbbottDeser,Woolgar:1994ar}, which guarantee the positivity of energy for an entire asymptotically AdS spacetime (defined by (\ref{boundary_hamiltonian}) with $\xi$ taken to coincide with the boundary time at the AdS boundary) assuming certain energy conditions. In our case we see that each physical spacetime must satisfy an infinite number of energy constraints, one positivity condition and a family of monotonicity conditions (discussed further below) for each subsystem $\Sigma_B$ associated with a boundary ball $B$.

The assumptions behind the theorems are also rather different. Typically, one requires that the matter in the theory is physically reasonable by assuming an energy condition\footnote{In \cite{AbbottDeser}, this was the dominant energy condition, while in \cite{Woolgar:1994ar}, a weaker averaged null energy condition was assumed.}, but there is no attempt to prove the energy condition from some underlying complete quantum theory. For our results, we assume that the spacetime arises in some consistent theory of quantum gravity with a CFT dual for which the holographic entanglement entropy formula (\ref{RT}) holds. Plausibly, this should be true for any consistent theory of quantum gravity whose low-energy equations of motion are Einstein's equations with couplings to arbitrary matter, so long as these couplings do not involve spacetime curvatures.\footnote{As we discuss further in section 5, we expect the result to hold also for more general theories of gravity, with an appropriately modified definition of the gravitational energy.}

For the global energy of an asymptotically AdS spacetime, positivity follows via AdS/CFT from the positivity of vacuum-subtracted energies in the CFT.\footnote{Alternatively, it can be shown based on causality in the CFT \cite{Page:2002xn}.} But the usual energy theorems show this positivity directly in general relativity by assuming an energy condition. In a similar way, while we have shown the energy and monotonicity results starting from properties of relative entropy in the CFT, it may be possible to prove these statements directly in general relativity by assuming some energy condition.\footnote{In general, this would only establish the energy condition as a sufficient condition for our (necessary) positive energy theorem.} This is an interesting problem for future work.

\subsection{Constraints on geometries}

The energy constraints that we have described may be viewed as purely geometrical constraints on the spacetimes that describe the entanglement entropies of consistent CFT states. Even when matter fields are present (without curvature couplings), the quantities appearing in the expressions (\ref{holographic_one}) dual to relative entropy depend only on the geometry. Certain asymptotically AdS geometries satisfy the constraints associated with positivity and monotonicity of relative entropy, while others violate them, and cannot correspond to consistent CFT states.

In assessing which geometries satisfy the constraints, we can work directly from the expressions in (\ref{holographic_one}) which are integrals over the codimension-two surfaces $B$ and $\tilde{B}$. Alternatively, we can rewrite the energy as a bulk expression, as in (\ref{Hamiltonian}). In order to make clear which constraints arise directly from the holographic entanglement entropy formula together with relative entropy inequalities without assuming the equations of motion, we can use the off-shell version of (\ref{JdQ}) \cite{hollands2013stability}
\be
\label{JCdQ}
J^{grav}_\xi =  d Q_\xi + C^{grav}_\xi,
\ee
where $Q_\xi$ is given in (\ref{defQ}), $C$ is defined in terms of the Einstein tensor $E_{ab}$ as
\be
\label{defC}
C^{grav}_\xi = {1 \over 8 \pi G_N} \xi^a E_a {}^b \epsilon_b \; .
\ee
and $J$ is given in equation (\ref{Puredefs}) of the appendix. Here, we are using the superscript `grav' to indicate that we are not considering the matter contributions to these quantities. Since the result (\ref{JCdQ}) is true off shell, it holds in general whether or not there are matter fields in the theory.
Applying the identity (\ref{JCdQ}) to (\ref{Hdiff}) with the definition (\ref{boundary_hamiltonian}), we can then write a bulk expression for relative entropy as
\be
 \label{bulkS}
S(\rho_B||\sigma_B) = \Delta \int_\Sigma (J_\xi - C_\xi) - \Delta\int_B \xi \cdot K,
\ee
with the boundary term vanishing when $B$ is regularized as a constant $z$ surface in Fefferman-Graham coordinates. Here, we can think of the first term involving $J_\xi$ as a gravitational contribution to the energy and the second term involving $C_\xi$ as a matter contribution to the energy, since on shell we can replace $E_{ab}$ appearing in $C_\xi$ with the matter stress tensor $T_{ab}$.

\subsection{General constraints from monotonicity}

In this section, we describe a minimal set of constraints on an asymptotically AdS spacetime $M$ which guarantee that all constraints associated with positivity and monotonicity of relative entropy for ball-shaped regions in the dual CFT will be satisfied.

\subsubsection*{A basis of constraints}

We note first that positivity of relative entropy for a region $B$ is equivalent to monotonicity applied to the case where the larger region is $B$ and the smaller region is the empty set (considered as a subset of $B$). Thus, it is sufficient to focus on the monotonicity constraint.

\begin{figure}
\centering
\includegraphics[width=0.5 \textwidth]{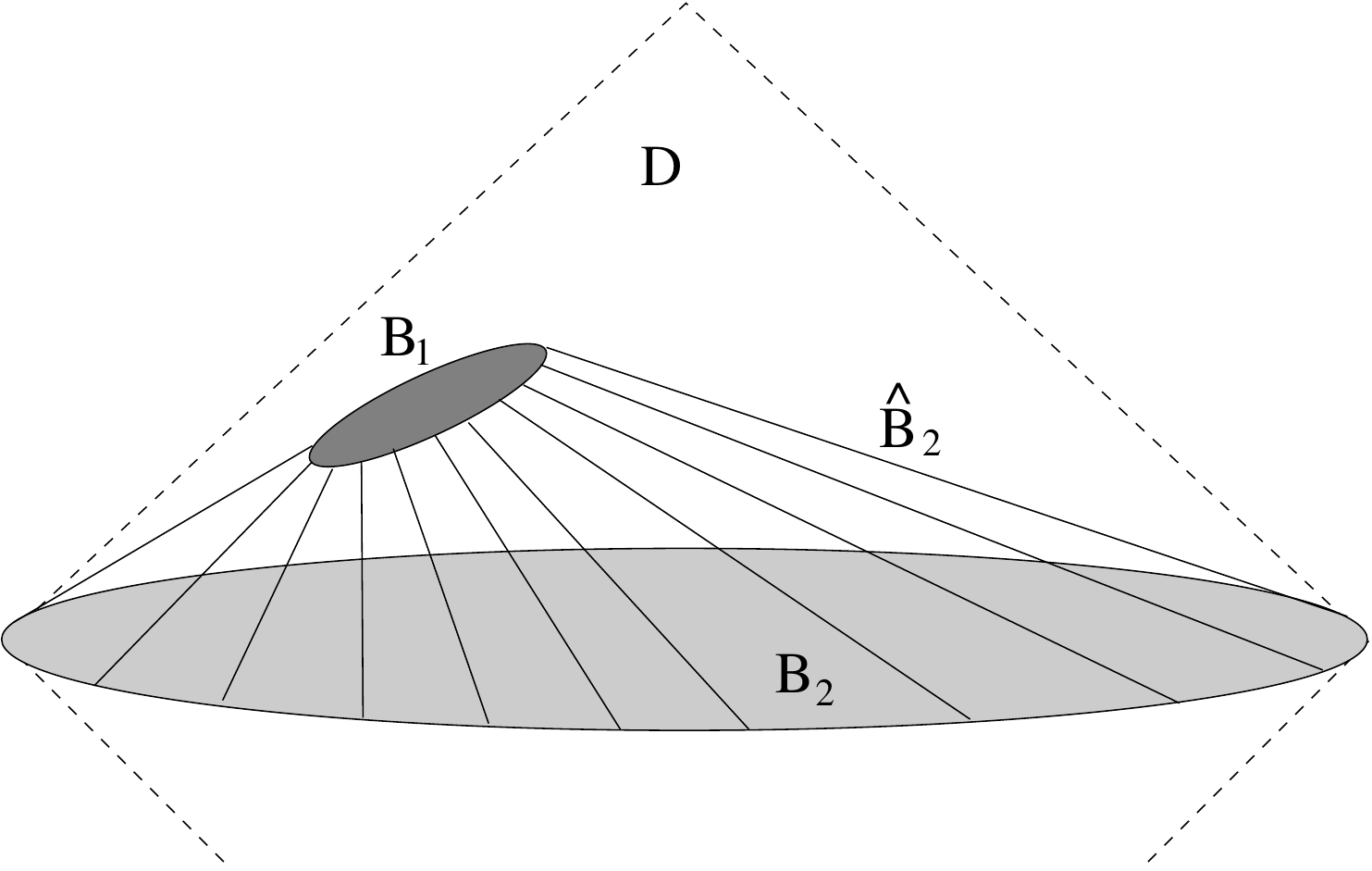}
\caption{For ball-shaped region $B_1$ in the domain of dependence $D$ of ball-shaped region $B_2$, monotonicity of relative entropy implies that the relative entropy for $B_2$ must be larger than or equal to the relative entropy associated with the subsystem $B_1$. Here, the surface $\hat{B}_2$ includes the ball $B_1$ and is a Cauchy surface for the same domain of dependence region $D$ as $B_2$, so it has the same relative entropy as for $B_2$.}
\label{monotonicityfig}
\end{figure}

For a relativistic conformal field theory, the monotonicity constraint
\be
\label{monotonicityB}
S(\rho^\Psi_{B_1}||\rho^{vac}_{B_1}) \le S(\rho^\Psi_{B_2}||\rho^{vac}_{B_2})
\ee
must hold for any two balls $B_1$ and $B_2$ for which the domain of dependence of $B_1$ is contained in the domain of dependence of $B_2$, as in figure \ref{monotonicityfig}, since in this case the fields on $B_1$ can be understood as a subset of the degrees of freedom associated with $B_2$.\footnote{To see this, we note first that the monotonicity constraint must hold for regions $A \subset B$ in any spatial slice. Considering a spatial slice that contains $B_1$ and $\partial B_2$ (possible since $B_1$ is in the domain of dependence of $B_2$), we have a monotonicity constraint associated with the regions $B_1$ and $\hat{B}_2$, where $\hat{B}_2$ is the region inside $\partial B_2$ on our spatial slice (see figure \ref{monotonicityfig}). But $\hat{B}_2$ and $B_2$ are just two different Cauchy surfaces for the same domain of dependence region. Thus, the corresponding density matrices are related by a unitary transformation, and the relative entropy associated with $\hat{B}_2$ is the same as the relative entropy associated with $B_2$. Thus, we can express the monotonicity constraint directly in terms of $B_2$ as in (\ref{monotonicityB}).}

For any $B_1$ and $B_2$ as above, there will be a one-parameter family of balls $B(\lambda)$ with $B(0) = B_1$, $B(1) = B_2$, and $B(\lambda_1)$ contained in the domain of dependence of $B(\lambda_2)$ for $\lambda_1 \le \lambda_2$. Applying the monotonicity constraint to any two infinitesimally nearby balls in this family, we obtain
\be
\label{moninf}
{d \over d \lambda} S( \rho^\Psi_{B(\lambda)}||\rho^{vac}_{B(\lambda)}) \ge 0 \, .
\ee
The collection of these infinitesimal conditions implies the finite constraint (\ref{monotonicityB}) upon integration over $\lambda \in [0,1]$. Thus, all relative entropy constraints for ball-shaped regions may be obtained from infinitesimal constraints (\ref{moninf}) associated with a ball $B$ and perturbations $B(\lambda)$ that enlarge the domain of dependence region.

\begin{figure}
\centering
\includegraphics[width=0.4 \textwidth]{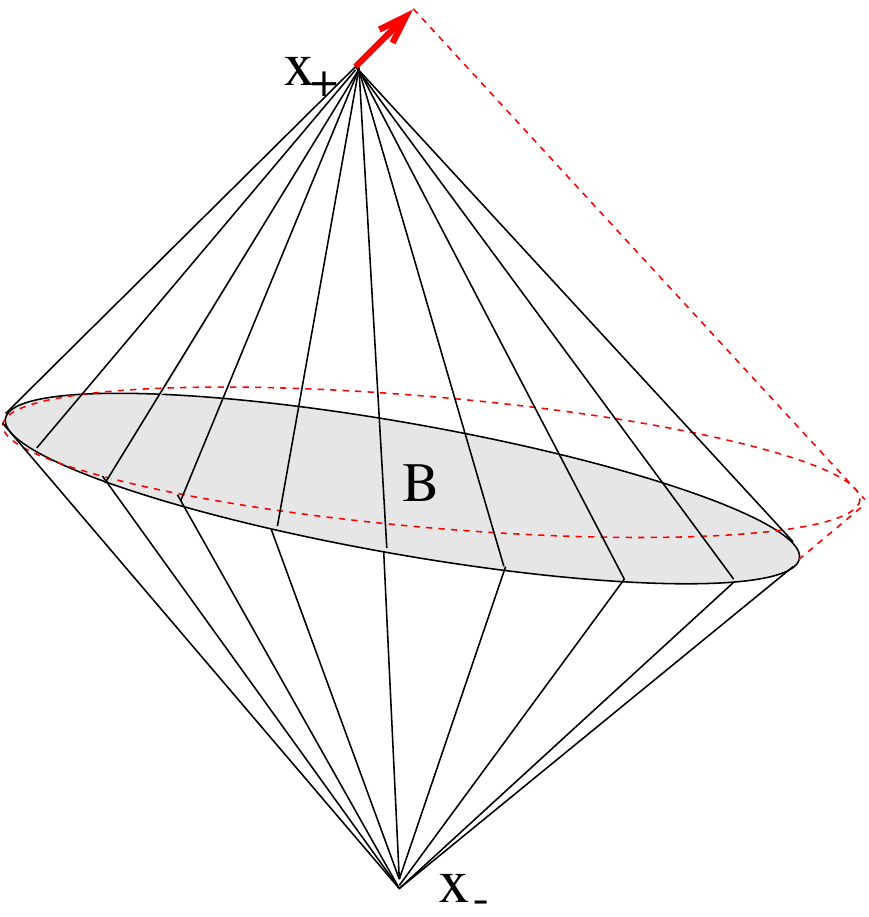}
\caption{One-to-one correspondence between balls $B$ and pairs of points $(x_+,x_-)$ with $x_+$ in the future of $x_-$. A minimal set of monotonicity constraints is obtained by considering deformations of the ball associated with shifting $x_+$ in a future lightlike direction (red arrow) or $x_-$ in a past lightlike direction. The boundary vector field $\Delta$ generates a conformal transformation that reverses this deformation.}
\label{xplusxminus}
\end{figure}

To describe these explicitly, we note that there is a one-to-one correspondence between balls $B$ and pairs $(x_-,x_+)$ of points with $x_+$ in the future of $x_-$, such that the boundary of the ball is the intersection of the future light cone of $x_-$ and the past light cone of $x_+$, as shown in figure \ref{xplusxminus}. Ball-enlarging transformations correspond to deformations which move $x_+$ in a future timelike direction and $x_-$ in a past timelike direction. To obtain the minimal set of constraints, it is enough to focus on a basis of such transformations: those that take either $x_+$ in a future lightlike direction with $x_-$ fixed or $x_-$ in a past lightlike direction with $x_+$ fixed. These correspond to infinitesimal perturbations that fix one point on the ball and translate the diametrically opposite point in a lightlike direction, as shown in  figure (\ref{xplusxminus}).

Each of these infinitesimal enlargements can be associated with a conformal transformation. Consider a ball of radius $R$ with center $x^\mu_0$ orthogonal to the timelike unit vector $u^\mu$. For this ball, $x_\pm = x_0 \pm R u$. Let $n_\mu$ be a spacelike unit vector orthogonal to $u^\mu$. Then $x_0 \pm n R$ are diametrically opposite points on the ball. A conformal transformation that holds $x_0 -n R$ fixed and moves $x_0 + n R$ in the positive/negative lightlike direction $n \pm u$ is given by $x^\mu \to x^\mu - \Delta^\mu$, where
\bea
\label{defDelta}
\Delta^\mu &=& \alpha x^\mu + \omega^{\mu} {}_\nu x^\nu + a^\mu \,, \\
\alpha &=& -{1 \over 2 R} \,,\cr
\omega^{\mu \nu} &=& \pm {1 \over 2 R} (n^\mu u^\nu - u^\mu n^\nu) \,,\cr
a^\mu &=& {1 \over 2 R} x^\mu_0 \mp {1 \over 2}(1 - {n \cdot x_0 \over R}) u^\mu - {1 \over 2} (1 \pm {u \cdot x_0 \over R}) n^\mu \nonumber\, .
\eea
In summary, we can define a basis of monotonicity constraints that are in one-to-one correspondence with pairs $(B,\Delta)$, where $B$ is a ball and $\Delta$ is an infinitesimal conformal transformation of this form.

\subsubsection*{Explicit geometrical constraints from monotonicity}

To describe the infinitesimal monotonicity constraints explicitly, it is useful to express  (\ref{moninf}) in a different way such that the ball remains fixed under the variation while the state changes. Given $B(\lambda)$, we define conformal transformations $U(\lambda)$ on the CFT associated to a family of conformal transformations that take $B(\lambda)$ back to the original ball $B(0)$. Then
\[
S( \rho^\Psi_{B(\lambda)}||\rho^{vac}_{B(\lambda)}) = S( \rho^{U(\lambda) \Psi}_{B}||\rho^{vac}_{B})\,,
\]
so the monotonicity constraint translates to
\be
\label{moninf2}
{d \over d \lambda} S( \rho^{U(\lambda) \Psi}_{B}||\rho^{vac}_{B})|_{\lambda=0} \ge 0 \, .
\ee
For our basis of transformations, we choose $U(\lambda)$ to be an infinitesimal transformation associated with generator $H_\Delta = -i \partial_\lambda U|_{\lambda=0}$ where $\Delta^\mu$ is any vector field of the form (\ref{defDelta}).

In the form (\ref{moninf2}), it is straightforward to translate the monotonicity constraint to an explicit constraint on geometries, given our result (\ref{Hdiff}). On the gravity side, the infinitesimal conformal transformation associated with $\Delta^\mu$ corresponds to a infinitesimal diffeomorphism
\be
\label{metic_change}
g \to g + {\cal L}_{\hat{\Delta}} g
\ee
 for some $\hat{\Delta}$ that extends $\Delta$ into the bulk. For an asymptotically $AdS$ spacetime in Fefferman-Graham coordinates, this vector field can be related explicitly to the boundary vector field $\Delta^a$ as
\be
\label{defDelta_bulk}
\hat{\Delta}^a(z,x) = (\hat{\Delta}^\mu(z,x),\hat{\Delta}^z(z,x)) = (\Delta^\mu(x), -\alpha z) \, .
\ee
where $\alpha$ is defined in (\ref{defDelta}). Since the relative entropy for ball $B$ is related to the gravitational Hamiltonian $H_{\xi_B}$ by (\ref{Hdiff}), and since the change in this Hamiltonian under a general variation of the metric is given by (\ref{varH}), we can immediately translate (\ref{moninf2}) to
\be
\label{monfinal}
\delta_{\hat{\Delta}} H_{\xi_B} = W({\cal L}_{\hat{\Delta}} g,{\cal L}_{\xi_B} g) \equiv \int_{\Sigma_B} \omega({\cal L}_{\hat{\Delta}} g,{\cal L}_{\xi_B} g) \ge 0\,,
\ee
where we recall that $W$ defines the symplectic form on the gravitational phase space associated with $\Sigma_B$. This gives an elegant gravitational interpretation of the general monotonicity constraint associated with the pair $(B,\Delta)$.

The result (\ref{monfinal}) is true on-shell. We can also obtain an off-shell version, starting from the result
\be\label{chi}
\delta S(\rho_B(\lambda)||\rho_B^{vac}) = \int_{\Sigma_B} {\rm d}\left[\delta Q_\xi(g) - \xi \cdot \theta(g, \delta g)\right]\,.
\ee
which follows from (\ref{Hdiff}) using the definitions (\ref{boundary_hamiltonian}) and (\ref{integrate_theta}). We will apply this to the metric perturbation defined by $\hat{\Delta}$. To proceed, we make use of the basic identity \cite{hollands2013stability}
\be
\label{identity}
{\rm d}\left[\delta Q_\xi(g) - \xi \cdot \theta(g, \delta g)\right] = \omega(g, \delta g,{\cal L}_\xi g) + \xi \cdot ( E(g) \cdot \delta g)- \delta  C_\xi(g)
\ee
where $E \cdot \delta g$ is defined to be the equations of motion term appearing in (\ref{eom}), and $C$ is defined by (\ref{JCdQ}). This identity holds off-shell for any fixed vector field $\xi$ and any variation of the metric, and is true for quantities $\omega$, $E$, $C$, $Q$, and $\theta$ defined with respect to any gravitational Lagrangian. Both $E$ and $C$ vanish if the equations of motion associated with this Lagrangian are satisfied. Applying (\ref{identity}) to (\ref{chi}) for the variation $g \to g + {\cal L}_{\hat{\Delta}} g$ in the case where the various quantities are defined with respect to the full Lagrangian of our theory including matter, and assuming that the equations of motion are satisfied, we immediately recover (\ref{monfinal}).

On the other hand, we can apply (\ref{identity}) off-shell to (\ref{moninf2}) in the case where the various quantities are defined with respect to the pure Einstein Lagrangian. In this case, we obtain the off-shell result
\be
\delta_{\hat{\Delta}} H_{\xi_B} = \int_{\Sigma_B} \omega^{grav}(g,{\cal L}_{\hat{\Delta}} g,{\cal L}_{\xi_B} g) + \xi \cdot ( E(g) \cdot {\cal L}_{\hat{\Delta}} g)- {\cal L}_{\hat{\Delta}} C_\xi(g) \ge 0
\ee
or, more explicitly,
\begin{eqnarray}
\label{gravconstraint4}
& &\int_\Sigma \epsilon_a\left\{\omega^a (g,{\cal L}_{\hat{\Delta}} g,{\cal L}_{X} g) + X^a E^{b c} ({\cal L}_{\hat{\Delta}} g)_{bc} - X^c E_c {}^{a} ({\cal L}_{\hat{\Delta}} g)_b {}^b \right.\\
& &\left. + 2 X^c E_{c b} ({\cal L}_{\hat{\Delta}} g)^{ba}- 2 X^c ({\cal L}_{\hat{\Delta}} E)_c {}^a \right\} \ge 0 \nn \, ,
\end{eqnarray}
where $E$ is the Einstein tensor and the tensor $\omega^a$ is given explicitly in appendix \ref{Waldappendix}. This provides a purely geometrical off-shell constraint that must hold for any consistent spacetime geometry.

We can obtain an alternative on-shell formula by replacing Einstein tensor with the matter stress tensor using the equations of motion
\be
\label{EE}
E^g_{ab} = {1 \over 2} T_{ab} \, .
\ee
With this replacement, we can think of the first term in (\ref{gravconstraint4}) as the gravitational contribution to (\ref{monfinal}) and the remaining terms as a matter contribution, which involves only the matter stress tensor. In this form, the constraint is something like an energy condition constraining the matter stress tensor. We will see specific examples below.

\subsection{Perturbative constraints}

We now consider spacetimes that are close to pure AdS and derive constraints on the geometries that follow from our general constraints above.

\subsubsection*{Review of perturbative implications of positivity}

Gravitational implications of the positivity of relative entropy in perturbation theory around the CFT vacuum were previously studied in \cite{Lashkari:2013koa, Lin:2014hva, Lashkari:2014kda, Lashkari:2015hha}; we briefly review these results and explain how to recover them from the positivity of our general formula \eqref{RelEnt}.

Since relative entropy vanishes for the reference vacuum state and is positive everywhere else, the first order variation of relative entropy vanishes. Combining the differential version (\ref{chi}) of our result with (\ref{identity}), using that $E(g) = {\cal L}_\xi g = 0$ for the background metric, and using that for pure gravity
\be
\label{defC}
C_\xi = {1 \over 8 \pi G_N} \xi^a E_a {}^b \epsilon_b \; ,
\ee
we obtain
\[
\int_{\Sigma_B} \xi^a \delta E_{ab} \epsilon^b = 0 \; .
\]
From the collection of these constraints for all $B$, it follows that $\delta E_{ab} = 0$ everywhere, i.e. that the first order perturbations to the geometry must satisfy Einstein's equations to linear order about AdS, as argued originally in \cite{Lashkari:2013koa,faulkner2014gravitation}.

To obtain the second order results from positivity of relative entropy, we can again start with the differential formula (\ref{chi}), replacing the integrand with the right side of the identity (\ref{identity}). Taking a second variation, we find \cite{Lashkari:2015hha}
\be\label{prel2}
\frac{d^2}{d\lambda^2}S(\rho(\lambda)||\sigma)|_{\lambda=0} = W_\Sigma(g, \gamma, \mathcal{L}_{\xi_B} \gamma)
\ee
where $\gamma = dg/d\lambda|_{\lambda=0}$ and $\xi_B$ is the bulk Killing vector in the AdS-Rindler wedge. The right hand side is defined in the general relativity literature as ``canonical energy" $\mathcal{E}(\gamma, \gamma)$ \cite{Hollands:2012sf}. Its positivity around a stationary black hole background implies linearized stability for axisymmetric perturbations to the black hole. Hence our result implies linearized stability of the AdS-Rindler wedge for physical perturbations in a theory of quantum gravity.

As explained in \cite{Lashkari:2015hha} (see \cite{Lin:2014hva, Lashkari:2014kda} for earlier related results), the positivity of the relative entropy at second order around the vacuum \eqref{prel2} can be massaged into a form resembling a manifest energy condition. Namely, {if one assumes the Einstein equations},
 one can write the canonical energy $\mathcal{E}$ as %
 \be
 \mathcal{E}(\gamma, \gamma) = -\int_\Sigma \xi^a(T_{ab}^{(2)} + T^{grav(2)}_{ab}){\epsilon}^b + {\rm boundary\,\,term}
 \ee
 where $T_{ab}^{(2)}$ are the terms in the matter stress tensor for bulk fields in AdS at second order in $\lambda$, and $T_{ab}^{grav(2)}$ is the expression quadratic in the first order metric perturbation that sources the next correction to the bulk metric when one perturbatively solves the Einstein equations. Up to the boundary term, this is the perturbatively corrected Rindler energy associated with the Killing vector $\xi_B$.

\subsubsection*{Perturbative implication of monotonicity}

Starting from (\ref{gravconstraint4}), we now derive the general constraints at second order coming from monotonicity of relative entropy. For a metric defined perturbatively as
\[
g(\mu) = g_0(0) + \mu g_1(0) + \mu^2 g_2(0) + \cdots
\]
the first new constraints from (\ref{gravconstraint4}) come at order $\mu^2$. These give
\be
\label{secondorder}
\delta_{\hat{\Delta}} H_{\xi_B}|_{{\cal O}(\mu^2)}  = \int_\Sigma \epsilon_a \left\{\omega^a(g_0,{\cal L}_{\hat{\Delta}} g_1,{\cal L}_{\hat{\zeta}_B} g_1) - 2 \xi_B^c ({\cal L}_{\hat{\Delta}} E^{(2)})_c {}^a \right\} \ge 0 \, ,
\ee
where $E^{(2)}$ represents the terms in the gravitational equations at second order in $\mu$.

We can compare this with the second order constraints due to positivity of relative entropy, which give (off-shell)
\be
\label{Fisher}
\int_\Sigma \epsilon_a \left\{ \omega^a(g_0, g_1,{\cal L}_{\xi_B} g_1) - 2 \xi_B^c ( E^{(2)})_c {}^a \right\} \ge 0 \, ,
\ee
As discussed above, the monotonicity constraints (\ref{secondorder}) must imply the positivity constraints (\ref{Fisher}), but in this case, we will see that they are stronger.

Using the explicit form (\ref{defxi}) of the bulk Killing vector $\xi_B$ in AdS and the expressions   (\ref{defDelta_bulk}) and (\ref{defDelta}) for $\hat{\Delta}$, we can give a more explicit formula for the second term in (\ref{secondorder}). We take a ball centered at $x_0$ with radius $R$ in a spatial slice perpendicular to a unit timelike vector $u$. We consider a deformation that holds a point $x^\mu_0 - n^\mu R$ on $\partial \tilde{B}$ fixed while shifting $x^\mu_0 + n^\mu R$ in the lightlike $n \pm u$ direction perpendicular to $\partial \tilde{B}$. Then the second term in (\ref{secondorder}) becomes
\[
- 2\int_\Sigma \epsilon_a \xi_B^c ({\cal L}_{\hat{\Delta}} E^{(2)})_c {}^a = {\pi \over R^2} \int_\Sigma \epsilon_\Sigma z \, d_{\tilde{B}}^2 \,  [{1 \over 2} (n \cdot x+ R)\partial_\pm E^{(2)}_{uu} \pm E^{(2)}_{u \pm} + {1 \over 2} z \partial_z E^{(2)}_{uu}] \, .
\]
where we have defined
\[
d_{\tilde{B}}^2 = R^2 - z^2 - (\vec{x}- x_0)^2 + (t - t_0)^2 \, ,
\]
and the integral runs over the bulk surface $\Sigma$ perpendicular to $(u^\mu,u^z=0)$ bounded by $B$ and $\tilde{B}$.

In \cite{Lashkari:2014kda}, monotonicity of relative entropy was used to derive constraints on the asymptotic metric of translation and time-translation invariant asymptotically $AdS^3$ spacetimes. Using the general result (\ref{secondorder}) above, we have checked that the constraints are precisely reproduced.

  \begin{figure}[h]
\centering
\includegraphics[width=0.7\textwidth]{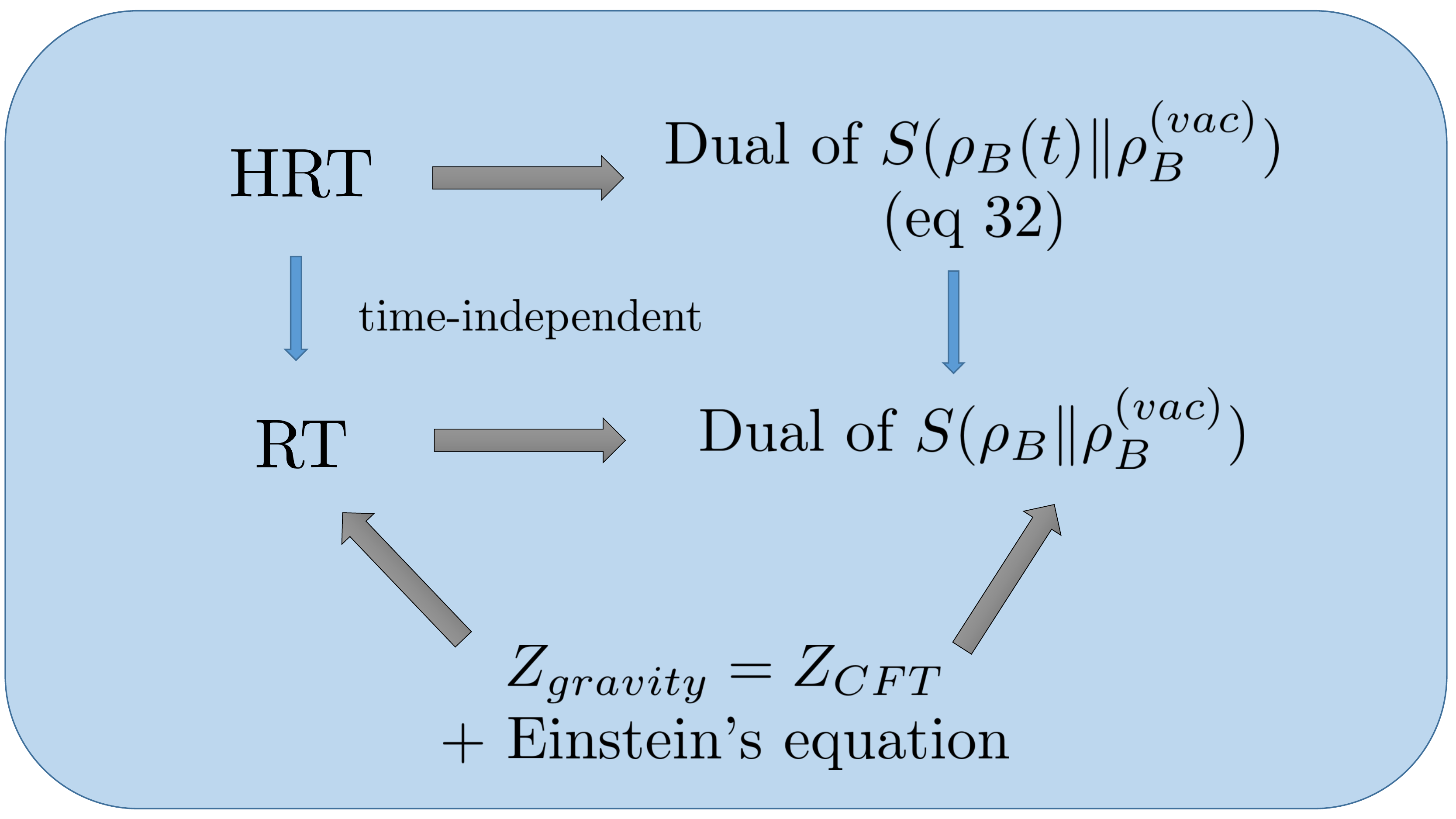}\\
\caption{\small{Bulk constraints and entries in the holographic dictionary: gray arrows represent proofs. They start from assumptions and point to conclusions. Blue vertical arrows signify restricting to special cases. HRT and RT stand for Hubeny-Rangamani-Takayanagi and Ryu-Takayanagi conjectures, respectively.}}
\label{fig6}
\end{figure}

\begin{figure}[h]
\centering
\includegraphics[width=0.7\textwidth]{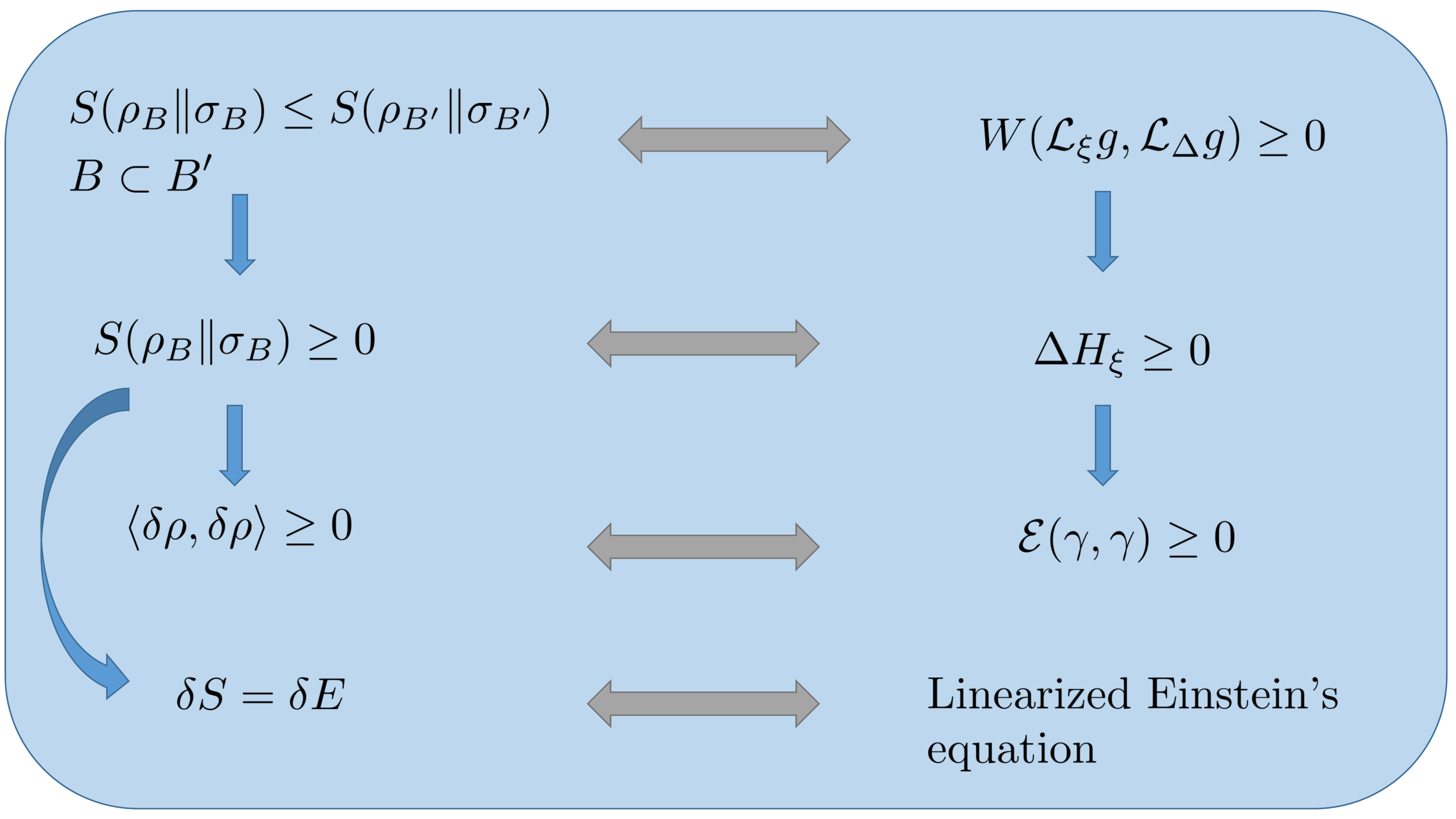}\\
\caption{\small{Information inequalities and bulk constraints.}}
\label{fig7}
\end{figure}

The relations between bulk constraints and information equalities and inequalities are summarized in figures \ref{fig6} and \ref{fig7}.

\section{Generalized Radon Transform}

In \cite{Lin:2014hva}, three of the authors of this paper
studied the holographic expression for
the relative entropy in the limit where the radius of the entanglement domain $B$ is
small. When the gravitational solution is time-reflection symmetric so that the
Ryu-Takayanagi formula can be used, they found,
\be
\label{AdSRadon}
\left( \frac{d^2}{dR^2} + \frac{1}{R}\frac{d}{dR}
- \frac{1}{R^2} \right) S(\rho_B || \rho_B^{{\rm vac}})
= 16 \pi^2 G_N \int_{\widetilde{B}} \varepsilon \sqrt{g_{\widetilde{B}}},
\ee
where $\varepsilon$ is the energy density of matter fields in the bulk
and $\sqrt{g_{\widetilde{B}}}$ is the induced volume form on $\widetilde{B}$.
In this limit, backreaction to the metric can be ignored and the bulk geometry
remains pure AdS. It was pointed out in \cite{Lin:2014hva} that
the  right-hand side of (\ref{AdSRadon}) takes the form of the Radon transform
of $\varepsilon$ on the $d$-dimensional hyperbolic space. It is known that
the Radon transform is invertible on hyperbolic space \cite{Helgason:1959,Rubin:2002}
and we can express the energy density
$\varepsilon$ as a superposition of the relative entropies for a family of domains
on the boundary.
In this way, we are able to reconstruct the local data on $\varepsilon$
in the bulk from the entanglement data represented by the relative entropy
on the boundary.

The holographic formula for the relative entropy derived in this paper enables
us to generalize result \eqref{AdSRadon} for finite $R$. As in \cite{Lin:2014hva}, we restrict our analysis to spacetimes which have time-reflection symmetry, such that the Ryu-Takayanagi surface is embedded in the time-reflection slice. The reflection symmetry ensures that in a neighborhood of this slice the metric components satisfy
\be
\label{metricparity}
\partial_t g_{tt}= \partial_t g_{\alpha\beta}= \OO(t), \qquad g_{t\alpha} = g_{\alpha t} = \OO(t),
\ee
where the Greek indices run over the spatial directions.

We start by parametrizing the Ryu-Takayanagi surface ending on the boundary of the sphere of radius $R$ as an even function of $t$,
\be
  f(x^a) = R.
\ee
Its gradient vector field ${}^*df = g^{ab} (\partial_a f) \partial_b$
is orthogonal to the Ryu-Takayanagi surface since any vector field $u^a$ parallel to the surface obeys $g_{ab}\ u^a ({}^*d f)^b
= u^a \partial_a f = 0$.
Using this,
we define field $\xi$ as a 1-form,
\begin{equation}
 \xi(R) = -2\pi t \sqrt{-g_{tt}} \ \frac{  f d f }{R ||df||}  -
\left( R - \frac{f(x^a)^2}{R} \right)  \frac{\pi\sqrt{-g_{tt}}\, dt}{||df||},
\end{equation}%
where $||df|| = \sqrt{g^{ab} \partial_a f \partial_b f}$.

Let us show that
this vector field satisfies the boundary conditions (\ref{x3}) - (\ref{x2}).
To check (\ref{x3}), we note that, in the AdS limit, $f \rightarrow \sqrt{t^2 + z^2 + x^2}$ and $\sqrt{-g_{tt}} \ ||df|| \rightarrow 1$. Therefore,
$\xi$ (with raised indices) reduces to the Killing vector field $\xi$ in AdS defined in eq.\ (21) of \cite{Lin:2014hva} and to the conformal Killing vector $\xi_B$ on the boundary.

It is easy to show that $\xi$ vanishes at $f(x^a) = R$ and (\ref{x2}) is satisfied. Using eq. \eqref{metricparity}, it can be checked that on the Ryu-Takayanagi surface
\be
\nabla^a \xi^b - \nabla^b \xi^a = 4\pi n^{ab},
\ee
so eq. \eqref{x1} is satisfied. Thus, $\xi^a$ satisfies the boundary conditions on the Ryu-Takayanagi surface.

 Differentiating
   $\xi(R)$ with respect to $R$,
we obtain
\begin{equation}
  \left( \frac{d}{dR} + \frac{1}{R}\right) \xi(R) =   2\pi \tau,
\end{equation}%
where
\begin{equation}
\label{deftau}
  \tau = - \frac{\sqrt{-g_{tt}}}{ ||df||} dt.
\end{equation}%
It follows that
\begin{equation}
   \left( \frac{d}{dR} + \frac{1}{R}\right) J_\xi = 2\pi J_{\tau},
\end{equation}%
and
\begin{equation}
   \left( \frac{d}{dR} + \frac{1}{R}\right) \int_\Sigma J_\xi
= 2\pi \int_\Sigma J_{\tau} + \int_{\partial \Sigma} v\cdot J_\xi,
\end{equation}%
where $v$ is the vector defined in the above so that $g_{ab} \ v^a  ( {}^*df)^b = 1$.

Since the relative entropy $S$ is expressed as
$\int_\Sigma J_\xi - \int_{\partial \Sigma}
\xi \cdot K$
minus the contribution from the vacuum AdS,
\begin{align} \left( \frac{d}{dR} + \frac{1}{R}\right) S &= \Delta \lsb 2\pi
\int_{\Sigma} J_\tau  + \int_{\partial \Sigma} v \cdot J_\xi
-2\pi \int_{\partial \Sigma} \tau \cdot K - \int_{\partial \Sigma} v \cdot \ d (\xi \cdot K) \rsb \nonumber \\
&= 2\pi \Delta H_\tau + \Delta \int_{\partial \Sigma}
v \cdot \left( J_\xi  - d(\xi \cdot K) \right),\nonumber
\end{align}
where $H_\tau$ on the right-hand side is a quantity obtained with respect to the timelike vector~$\tau$ as
\begin{equation}
\label{Hamiltoniantau}
H_\tau = \int_{\Sigma} J_\tau
- \int_{\partial \Sigma} \tau \cdot K.
\end{equation}
The vector field $\tau$ is independent of $R$.\footnote{Note, however, that it still depends on the ball $B$ through the function $f$ appearing in (\ref{deftau}).}
In the AdS limit (with raised indices) it becomes $\tau \rightarrow \partial_t$.

Since the vector field $\tau$ does not vanish on the minimal surface,
the the Wald-Zoupas integrability condition \eqref{integrability} does not
necessarily hold. Thus, strictly speaking,
$H_\tau$ is not a quasi-local energy in the sense
defined in section 2.2. On the other hand, the formulas we will derive
below using $H_\tau$ give natural generalizations of the results
in \cite{Lin:2014hva} on the positivity and the Radon transform of the matter energy density.

Since $\xi$ vanishes on the Ryu-Takayanagi surface,
\begin{equation} d(\xi \cdot K) = {\cal L}_\xi K =
\theta ( {\cal L}_\xi g),
\end{equation}
so that (using definition \eqref{JefD} for $J_\xi$) we find
\begin{equation} J_\xi  - d(\xi \cdot K) = 0 \end{equation}
on the Ryu-Takayanagi surface.
Therefore, the $R$ derivative of the relative entropy can be expressed as
\begin{equation}  \left( \frac{d}{dR} + \frac{1}{R}\right) S = 2\pi \Delta H_\tau,
\end{equation}
where $H_\tau$ is given by eq. \eqref{Hamiltoniantau}.
The positivity and monotonicity of $S$ mean that $\Delta H_\tau$ is non-negative.

One more $R$ derivative gives
\begin{equation}
\label{equ87}
\frac{d}{dR}  \left( \frac{d}{dR} + \frac{1}{R}\right) S
= 2\pi \Delta \int_{\widetilde{B}} v \cdot \left( J_\tau - d(\tau \cdot K) \right).
\end{equation}
This generalizes the Radon transform formula (\ref{AdSRadon}) for finite $R$.
It would be interesting to determine if this can be inverted
to find an expression for the local quantity
$(J_\tau - d(\tau \cdot K))$ in the bulk from the entanglement data represented
by the relative entropy.

For a theory of gravity plus a scalar field,\footnote{The argument below goes through, essentially unchanged, for multiple scalar fields.}
\be
L = \frac{1}{16\pi G_N}R - \frac{1}{2}\lb\partial \phi\rb^2 - V(\phi),
\ee
the right-hand side of eq. \eqref{equ87} can be further simplified by using the identity (see e.g. eq. (34) of \cite{Iyer:1995kg})
\be
\label{thisuglything}
\delta \lb \KK\epsilon^{(d)} \rb = \frac {1}{2} \epsilon^{(d)} \left( \KK_{ab} - \gamma_{ab}\KK \right)\delta \gamma^{ab}
+ \frac {1}{2}\epsilon^{(d)}n^a\left(- \nabla^b\delta g_{ab} + g^{cd}\nabla_a\delta g_{cd}  \right),
\ee
which holds for arbitrary variations, where we have dropped a total derivative term. Here $\KK_{ab}$, $\gamma_{ab}$ and $\epsilon^{(d)}$ are the extrinsic curvature, induced metric and volume form on $\partial\MM$ embedded in the slice of time reflection symmetry, and $n^a$ is the spacelike unit normal to $\partial \MM$.

Defining the boundary term as\footnote{As explained in \cite{Iyer:1995kg}, we can add to $K$ any function $S_0$ that depends only on the intrinsic geometry of $\partial \MM$. Demanding to recover the modular Hamiltonian expectation value on the boundary fixes $S_0$ on $B$ as in eq. \eqref{Kbdyis}, however it does not determine $S_0$ on $\tilde B$.}
\be
\label{Kbdyis}
K = - \frac{1}{8\pi G_N} \lb \KK + \frac{d-1}{\ell} \rb \epsilon^{(d)} + F(\phi),
\ee
eq. \eqref{thisuglything} turns into
\be
\label{eq89}
\delta K = \theta \lb \delta g \rb + \frac {1}{2\kappa^2} \epsilon^{(d)} \left( \KK^{ab} - \gamma^{ab}\KK - \gamma^{ab} \frac{d-1}{\ell} \right)\delta \gamma_{ab} + \lb \nabla^a \phi \rb \delta \phi \, \epsilon_a - \delta F.
\ee
Here $\theta$ is defined for the full theory, and $F$ denotes any scalar field counterterms we may need to add to $K$ to recover the modular Hamiltonian on $B$.\footnote{Since we are mostly interested in normalizable  scalar fields, it should be fine to ignore the counterterms in most, if not all, situations.}

Eq. \eqref{Kbdyis} is an explicit construction for the boundary term $K$. It can be checked that with $K$ defined in this manner and $\xi$ as above, the difference in integrals of $Q_\xi - \xi \cdot K$ on $B$ and $\tilde B$ equals the difference in entanglement entropy and modular Hamiltonian expectation value, respectively.

For arbitrary variations, the last three terms on the right-hand side of eq. \eqref{eq89} do not vanish. However, for $\delta = \LL_\tau$, parity conditions \eqref{metricparity} (and the fact that $\phi$ is even under time reflections) ensure that in a neighborhood of the Ryu-Takayanagi surface these terms are of order $\OO(t)$. Thus, on the Ryu-Takayanagi surface we have
\be
\LL_\tau K = \theta\lb \LL_\tau g \rb.
\ee
This simplifies eq. \eqref{equ87} to
\be
\frac{d}{dR}  \left( \frac{d}{dR} + \frac{1}{R}\right) S
= -2\pi \Delta \int_{\widetilde{B}} v \cdot \tau\cdot \left( L - dK \right).
\ee
Thus, for pure gravity with normalizable scalar fields, an inversion formula for the Radon transform would reconstruct the bulk action from relative entropy.\footnote{Such a reconstruction, if it exists, should have a natural way of dealing with the ambiguities in the definition of $K$ on $\tilde B$.}

\section{Discussion}

In this paper we have seen that for holographic conformal field theories in which the Ryu-Takayanagi formula (and its covariant generalization) hold, relative entropy for a ball-shaped region $B$ in the CFT maps (at the classical level) to the vacuum-subtracted energy $H_\xi$ associated to a vector field $\xi$ that behaves like a ``local'' Killing vector near the AdS boundary and near the extremal surface $\tilde{B}$ where it vanishes.

We expect that a similar result holds for more general theories of gravity (e.g. including higher curvature terms). Starting from (\ref{eom}) with a more general gravitational Lagrangian, it is possible using the equations in that section to define quantities $\theta$, $\omega$, $J_\xi$,  $Q_\xi$, and $H_\xi$ related to the more general Lagrangian. To demonstrate an equivalence between relative entropy and $\Delta H_\xi$, it is necessary to show the analogue of equations (\ref{holographic_energy}) and (\ref{holographic_entropy}). Our argument for (\ref{holographic_energy}) goes through in the general case since the results in \cite{faulkner2014gravitation} apply generally. However, to show (\ref{holographic_entropy}), it is necessary to argue that the generalized holographic entanglement entropy functional (which is believed to equal the Wald functional for black hole entropies plus certain corrections depending on extrinsic curvatures) can be written as an integral over $Q_\xi$, with some suitable conditions on $\xi$ generalizing (\ref{x1}) and (\ref{x2}) and making use of the available freedom in the definition of $Q_\xi$.\footnote{We thank Rob Myers for a discussion on this point.} We leave this as a question for future work.

In this paper, we have focused on the leading large $N$ contribution to relative entropy, making use of the leading-order holographic entanglement entropy formula. According to \cite{Faulkner:2013ana}, the $1/N$ corrections to CFT entanglement entropy correspond to the entanglement entropy of bulk quantum fields across the extremal surface $\tilde{B}$ (made finite by the intrinsic regulator provided by quantum gravity). Including this additional term, our result becomes\footnote{As argued in \cite{swingle2014universality}, the holographic formula for the modular Hamiltonian variation does not require modification, provided we assume that the bulk matter stress tensor dies off sufficiently rapidly at the boundary.}
\be
\label{REboundarybulk}
S(\rho_B||\sigma_B) = \Delta H_{\xi_B} - \Delta S_{\Sigma_B} \; .
\ee
This is reminiscent of the CFT definition (\ref{REdiff}) of relative entropy. In the recent works \cite{Jafferis:2015del, Dong:2016eik}, it has been argued that at a perturbative level, CFT relative entropy  for a region $B$ to order $1/N$ maps over to semiclassical bulk relative entropy for the region $\Sigma_B$. For this equivalence to extend to the non-perturbative level that we have considered in this paper, it would be necessary to identify $\Delta H_{\xi_B}$ with the change in the expectation value of the bulk modular Hamiltonian associated with the AdS vacuum. At the semiclassical level, it was argued in \cite{swingle2014universality} that this modular Hamiltonian is given by
\[
H_\Sigma = \int_\Sigma \xi^a T_{ab} \epsilon^b \; ,
\]
where $T_{ab}$ includes contributions from all perturbative fields including the graviton and $\xi$ is the Killing vector (\ref{defxi}) associated with the region $\Sigma_B$ in AdS. If we conjecture that this operator is well-defined non-perturbatively and that its expectation value for general states gives the energy $H_{\xi_B}$, then it would follow that the boundary relative entropy  and bulk relative entropy can be identified even at the non-perturbative level (at least when the subsystems are ball-shaped and the reference state is the vacuum).

The results of this paper lend support to the idea of subregion duality in AdS/CFT. In quantum field theory, given a spatial region $A$, the set of fields and observables restricted to the associated domain of dependence region $D_A$ form a natural subsystem of the field theory, since such observables do not depend on the fields outside of the region $A$, and naturally form an algebra on their own. In a sense, the field theory on such a region $A$ is a self-contained physical system. For a holographic CFT, it is natural to ask (see e.g. \cite{Bousso:2012sj,Czech:2012be,Hubeny:2012wa}) whether such a system can be considered to have a gravity dual. The results in this paper provide further evidence that such a subsystem of the CFT describes the gravitational physics within the ``entanglement wedge'' of the CFT \cite{Czech:2012be}, the region between the boundary domain of dependence region $D_B$ and the extremal surface $\tilde{B}$. Specifically, we have found that it is possible to define a phase-space Hamiltonian $H_\xi$ associated with this region when $B$ is a ball-shaped boundary region, and argued that the value of this energy relative to the pure AdS vacuum state is always positive. Thus, for the class of entanglement wedge geometries corresponding to a given ball-shaped boundary region, it is possible to define self-contained dynamics associated with a positive-definite Hamiltonian.

\section*{Acknowledgments}

We thank Xi Dong, Thomas Faulkner, Simon Gentle, Daniel Harlow, Ken Intriligator, Lampros Lamprou, Aitor Lewkowycz, Hong Liu, Juan Maldacena,
Travis Maxfield, John McGreevy, Rob Myers, Ingmar Saberi, Jaewon Song, and Edward Witten for discussions. The research of MVR is supported in part by the Natural Sciences and Engineering Research Council of Canada, and by grant 376206 from the Simons Foundation.
The research of HO and BS are supported in part by
U.S.\ Department of Energy grant DE-SC0011632 and by Caltech's
Walter Burke Institute for Theoretical Physics and Moore Center for Theoretical Cosmology and Physics. The research of HO is
also supported in part by the Simons Investigator Award,
by the World Premier International Research Center Initiative (WPI Initiative),
MEXT, Japan,
by JSPS Grant-in-Aid for Scientific Research C-26400240, and
by JSPS Grant-in-Aid for Scientific Research on Innovative Areas
15H05895. NL is supported in part by funds provided by MIT-Skoltech Initiative.
JL acknowledges support from the Schmidt Fellowship and the U.S.
Department of Energy.
We thank the hospitality of the Institute for Advanced Study, where HO was Director's Visiting Professor in the fall 2015.
HO also thanks the hospitality of the Aspen Center for Physics,
the Simons Center for Geometry and Physics, and the Center for Mathematical
Sciences and Applications and the Center for the Fundamental Laws of Nature
at Harvard University, where he is a visiting scholar in the spring 2016. B.S. thanks MIT, Stanford University, and the Simons Center for Geometry
and Physics for hospitality.

\appendix

\section{Relative entropy as generalized free energy}\label{freeEnergy}
\begin{figure}[t]
\centering
\includegraphics[width=0.7\textwidth]{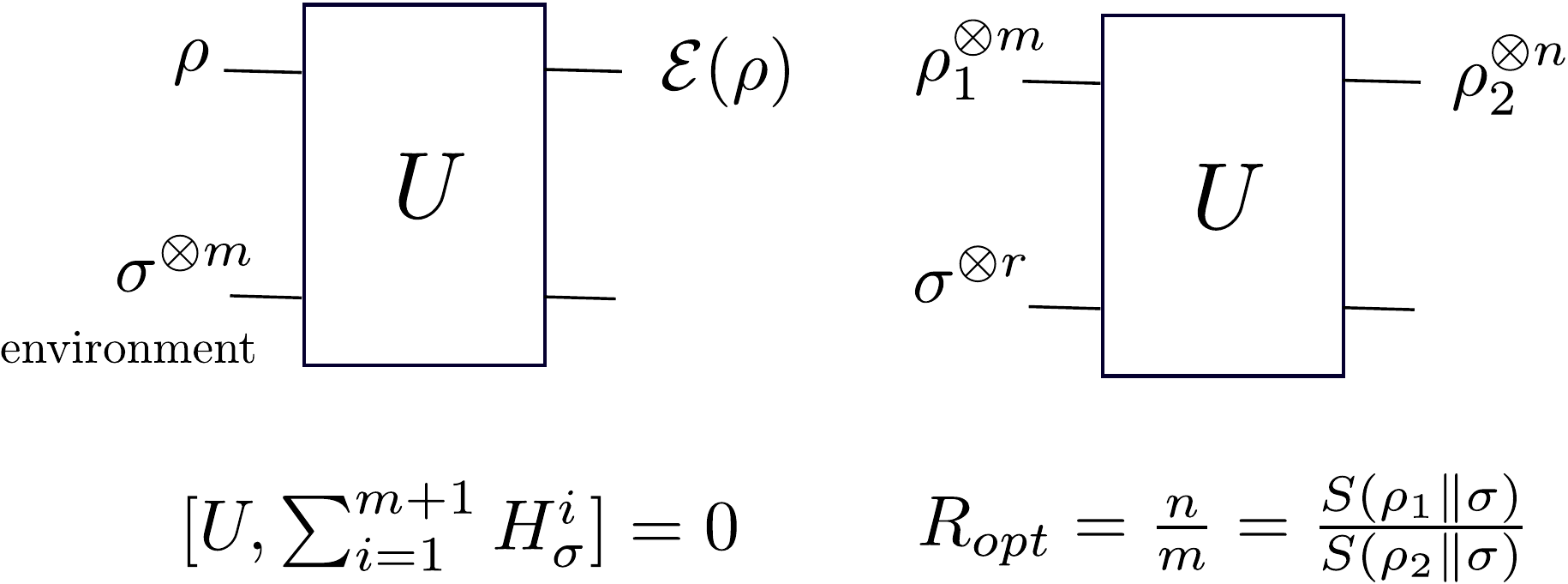}\\
\caption{\small{The set of allowed operations.}}
\label{fig5}
\end{figure}

Consider a quantum ``thermodynamic theory" (resource theory) in which $H_\sigma$ and $\sigma=e^{-H_\sigma}$ play the role of Hamiltonian and equilibrium state, respectively. In thermodynamics, we restrict the set of allowed operations to those that conserve the total energy of the system and environment combined. A natural generalization of this principle to our case is to define the set of allowed operations to be the unitaries that act on the system and arbitrary number of copies of the equilibrium state conserving the total ``energy"; see figure \ref{fig5}. In other words, the most general evolution is a quantum channel defined by
\bea\label{channel}
\mathcal{E}(\rho)=tr_{env}\left[U(\rho\otimes \sigma^{\otimes m}_{env}) U^\dagger\right],\qquad [\sum_{i=1}^{m+1} H^i_\sigma,U]=0.
\eea

In this framework, we are going to interpret relative entropy as the excess ``free energy" of $\rho$ from equilibrium,
\bea
S(\rho\|\sigma)&=&F_\sigma(\rho)-F_\sigma(\sigma),\nn\\
F_\sigma(\rho)&=&tr(\rho H_\sigma)-S(\rho).
\eea
Here we mention three important properties of relative entropy that makes this interpretation natural.

\begin{enumerate}
\item {\bf Equilibrium state minimizes free energy:} For any non-equilibrium state, one expects free energy to be larger than its equilibrium value. This is indeed true since relative entropy of any two states is non-negative and becomes zero if and only if the two states are the same.

\item {\bf Free energy is never created spontaneously:} The class of operations defined in (\ref{channel}) is a quantum channel. According to data-processing inequality, relative entropy is non-increasing in quantum channels \cite{uhlmann1977relative},
\bea\label{monotone}
S(\rho\|\sigma)\geq S(\mathcal{E}(\rho)\|\mathcal{E}(\sigma))=S(\mathcal{E}(\rho)\|\sigma).
\eea
Therefore, relative entropy quantifies a resource. It never increases spontaneously, and can only be distilled or diluted.

\item {\bf Free energy quantifies how much work (resources) can be extracted:} From (\ref{monotone}) we know that if we convert $m$ copies of low resource state $\rho_1$ to $n$ copies of resourceful states $\rho_2$, we always have the inequality $n S(\rho_2\|\sigma)\leq m S(\rho_1\|\sigma)$; figure \ref{fig5}. In other words, the optimal rate at which one can distill the resource is
\bea
R_{opt}(\rho_1\to\rho_2)=\frac{n}{m}=\frac{F_\sigma(\rho_1)-F_\sigma(\sigma)}{F_\sigma(\rho_2)-F_\sigma(\sigma)}.
\eea
This was shown in the context of generic resource theories in \cite{brandao2013resource}.

\end{enumerate}

\section{Forms}
\label{Waldappendix}
Here, we list explicit expressions for the various forms appearing in section 2.2 in the case of pure Einstein gravity with a cosmological constant.
To begin, we define the forms
\[
\boldsymbol{ \epsilon}_{c_1 \dots c_k} =  {1 \over (d-k+1)!} \sqrt{-g} \epsilon_{c_1 \dots c_k a_{k+1} \cdots a_{d+1}} dx^{a_{k+1}}\wedge \dots \wedge dx^{a_{d+1}} \; ,
\]
which provide volume forms for codimension $k$ submanifolds. For a general vector field $X$, we have \cite{hollands2013stability, faulkner2014gravitation}
\begin{eqnarray}
\label{Puredefs}
L &=& {1 \over 16 \pi G_N} R - \Lambda  \\
\theta &=& {1 \over 16 \pi G_N} \epsilon_a (g^{ac} g^{bd}-g^{ad} g^{bc}) \nabla_d {d \over d \lambda} g_{bc} \nn\\
E^g_{ab} &=&  R_{ab} - {1 \over 2} g_{ab} R + 8 \pi G_N g_{ab} \Lambda \nn\\
C_X &=& {1 \over 8 \pi G_N} X^a E^g_{ab} \epsilon^b \nn\\
Q_X &=& {1 \over 16 \pi G_N} \nabla^a X^b \epsilon_{ab} \label{defq}\nn\\
J_X &=& \frac{1}{8\pi G_N}\nabla_e\left( \nabla^{[e}X^{d]}\right)\epsilon_d +{1 \over 8 \pi} X^a E^g_{ab} \epsilon^b
\nn\\
\omega &=& {1 \over 16 \pi G_N} \epsilon_a P^{abcdef} (\gamma^2_{bc} \nabla_d \gamma^1_{ef} - \gamma^1_{bc} \nabla_d \gamma^2_{ef}) \nn\\
P^{abcdef} &=& g^{ae} g^{fb} g^{cd} - {1 \over 2} g^{ad} g^{be} g^{fc} - {1 \over 2} g^{ab} g^{cd} g^{ef} - {1 \over 2} g^{bc} g^{ae} g^{fd} + {1 \over 2} g^{bc} g^{ad} g^{ef} \nn\\
\delta Q_X - X \cdot \theta(g, \delta g) &=& {1 \over 16 \pi G_N} \epsilon_{ab} \left\{\gamma^{ac} \nabla_c X^b - {1 \over 2} \gamma_c{}^c \nabla^a X^b + \nabla^b \gamma^a {}_c X^c - \nabla_c \gamma^{ac} X^b + \nabla^a \gamma^c {}_c X^b \right\} \, .\nn
\end{eqnarray}

\section{Gaussian null coordinates and the vector field $X$}\label{a2}

An essential part of our discussion is the existence of a vector field $\xi$ which reduces to $\zeta_B$ at the AdS boundary and satisfies $\xi=0$ and $\nabla^{a} \xi^b  = 2 \pi n^{ab}$ on the surface $\tilde{B}$. In this appendix, we describe an explicit construction for this vector field near $\tilde{B}$, making use of Gaussian null coordinates.

To define the Gaussian null coordinates, we start with coordinates $x^i$ on our surface $\tilde{B}$, and consider a normal null vector field $N$ on the surface $\tilde{B}$ which generates the future-directed lightsheet in the direction toward the boundary. Parametrizing the geodesics generated by vectors $N^\mu$ by a parameter $u$, we can associate coordinates $(x^i,u)$ to a point $p$ on the lightsheet in a neighborhood of $\tilde{B}$ that lies at parameter value $u$ on the geodesic from the point at coordinates $x^i$. The assignment $(x^i,u)$ will be unambiguous for a sufficiently small neighborhood of $\tilde{B}$.

Finally, we consider the past-directed null vector field $L$ defined on the lightsheet such that $L \cdot \partial_u = 1$ and $L \cdot \partial_i = 0$. Introducing the affine parameter $r$ for the geodesics generated by $L$, we can now associate coordinates $(r,u,x^i)$ to any point Q in a neighborhood of $\tilde{B}$, where Q lies at parameter $r$ along the geodesic from the point P on the lightsheet with coordinates $(u,x^i)$. Again, this gives a unique specification of coordinates for points in a sufficiently small neighborhood of $\tilde{B}$. This defines a set of Gaussian null coordinates in the neighborhood of $\tilde{B}$.

In these coordinates, the metric takes the form (for a detailed argument, see section 2.1 of \cite{Kunduri:2013ana})
\[
ds^2 = 2 du dr + A(r,u,x^i) du^2 + B_i(r,u,x^i) du d x^i + C_{ij}(r,u,x^i) dx^i dx^j \, ,
\]
where $A$ and $B_i$ vanish for $r=0$. From this expression, it is straightforward to check that the vector field
\[
\xi = 2 \pi (u \partial_u - r \partial_r)
\]
satisfies the desired conditions, $\xi=0$ and $\nabla^{a} \xi^b  = 2 \pi n^{ab}$ on the surface $\tilde{B}$. Away from $\tilde{B}$, we are free to choose $\xi$ as we like in order to approach the boundary vector field $\zeta$.

 \section{Conformal map to hyperbolic coordinates}\label{conformaltrans}
\begin{figure}[t]
\centering
\includegraphics[width=0.6\textwidth]{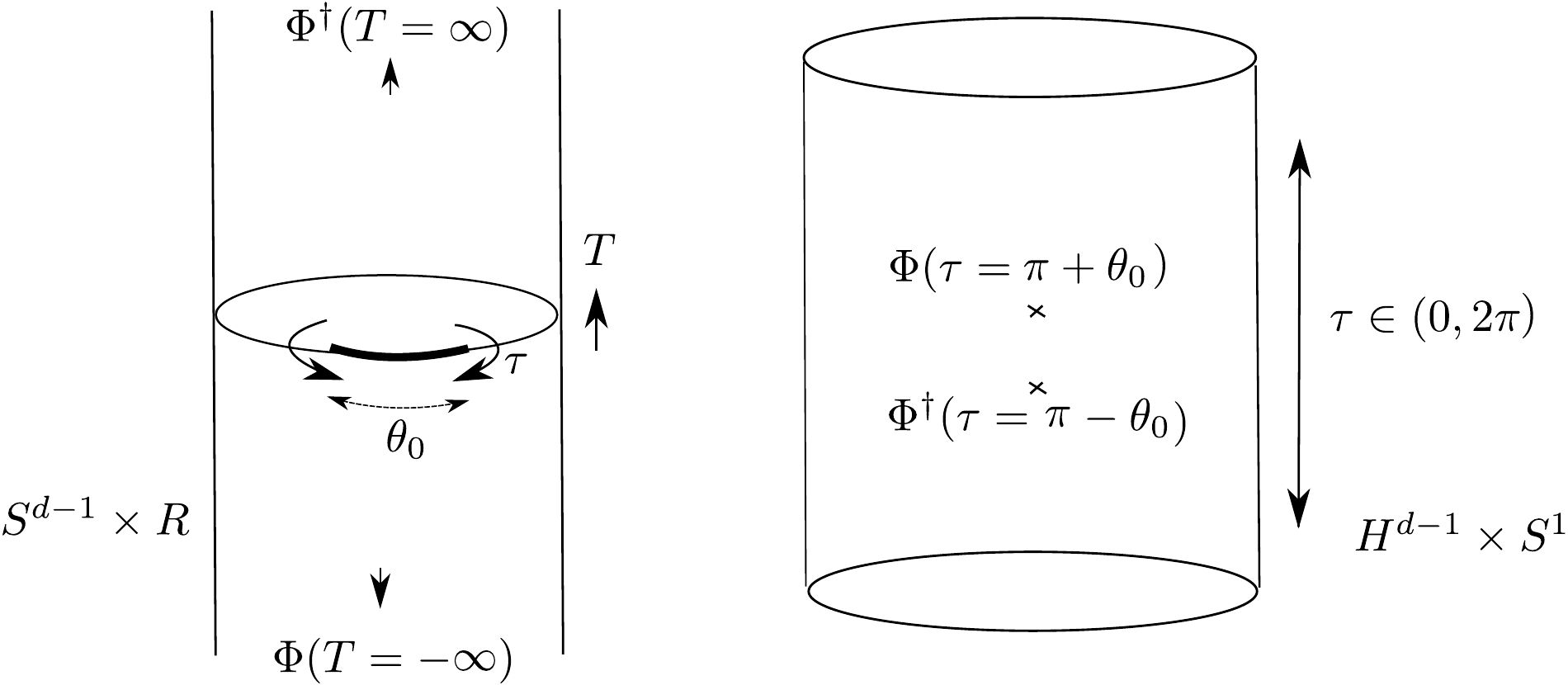}\\
\caption{\small{The conformal transformation from the ball to the hyperbolic plane.}}
\label{fig4}
\end{figure}

Consider a conformal field theory on a sphere $S^{d-1}$ of radius $R/\epsilon$ with $\epsilon\ll 1$ acting as an infrared regulator for the theory in flat space.
The partition function associated with an excited state is given by a Euclidean path-integral over cylinder $S^{d-1}\times R$ with operators $\Phi$ and $\Phi^\dagger$ that create the state inserted at $T=\pm\infty$. Here, $T$ parametrizes the Euclidean time along the cylinder. The metric is
\bea
ds^2=dT^2+\lb R/\epsilon\rb^2\lb d\theta^2+\sin^2\theta d\Omega_{d-2}^2\rb.
\eea
We make the following coordinate transformation
\bea
&&\tanh(T \ep/R)=\frac{\sin(\ep/R)\sin(\tau)}{\cosh u+\cos(\ep/R)\cos\tau},\nn\\
&&\tan\theta=\frac{\sin(\ep/R)\sinh u}{\cosh u\cos(\ep/R)+\cos\tau},
\eea
 that brings the metric to the form
 \bea
&&ds^2=\Omega^2\lb d\tau^2+(du^2+\sinh^2 ud\Omega_{d-2}^2)\rb,\nn\\
&&\Omega^2=\frac{R^2\sin^2(\ep/R)/\ep^2}{(\cosh u\cos(\ep/R)+\cos\tau)^2+\sin(\ep/R)^2\sinh^2u}.
 \eea
 A Weyl transformation eliminates the factor $\Omega^2$ leaving the metric on $H^{d-1}\times S^1$
 \bea
 ds^2=d\tau^2+(du^2+\sinh^2 ud\Omega_{d-2}^2).
 \eea
 The $\tau$ direction is the thermal circle with periodicity $2\pi$. The two balls $\theta\leq \ep$ at $T=0^\pm$ are mapped to the hyperbolic planes at $\tau=0$ and $\tau=2\pi$. The operator insertions at $r=0$ and $T=\pm\infty$ are respectively mapped to $u=0$ and $\tau=\pi\mp \ep$; see figure \ref{fig4}.

\providecommand{\href}[2]{#2}\begingroup\raggedright\endgroup

\end{document}